\documentclass[a4, 11pt]{article}
\usepackage[T1]{fontenc}
\usepackage{bm}
\usepackage[sort&compress,numbers]{natbib}
\usepackage{enumitem}
\usepackage{algorithm}
\usepackage{amssymb}        
\usepackage{latexsym}
\usepackage{algpseudocode}
\usepackage{graphicx}
\usepackage{verbatim}
\usepackage{hyperref}

\usepackage{tikz}
\usepackage{subcaption}

\usepackage{amssymb}        
\usepackage{latexsym}
\usepackage{times}
\usepackage{array}
\usepackage{subcaption}

\usepackage{graphicx}

\author{}

\date{}

\title{New Approximation Algorithms for Maximum Asymmetric Traveling Salesman and Shortest Superstring}

\newcommand{\dowod}{\noindent{\bf Proof.~}}
\newcommand{\koniec}{\hfill $\Box$\\[.1ex]}
\newtheorem{fact}{Fact}

\newcommand{\Kd}{{\cal K}}

\newcommand{\Kr}{{\cal K}_4}

\newtheorem{lemma}{Lemma}
\newtheorem{theorem}{Theorem}
\newtheorem{corollary}{Corollary}
\newtheorem{definition}{Definition}
\newtheorem{observation}{Observation}
\newtheorem{claim}{Claim}

\begin{document}

\maketitle
\thispagestyle{empty}
\begin{abstract}

In the maximum asymmetric traveling salesman problem (Max ATSP), we are given a complete directed graph with nonnegative weights on the edges, and we wish to compute a traveling salesman tour of maximum weight. 
In this paper, we present a fast combinatorial $\frac{7}{10}$-approximation algorithm for Max ATSP. It is based on techniques of  \emph{eliminating} and \emph{diluting} problematic subgraphs with the aid of \emph{half-edges} and a novel method of edge coloring. (Informally speaking, a \emph{half-edge} of an edge $(u,v)$ is ``either a head or a tail of $(u,v)$''.) The technique of \emph{diluting} a problematic subgraph $S$ consists in a seeming reduction of its weight, which allows its better handling.

The current best approximation algorithms for Max ATSP, achieving an approximation guarantee of $\frac{2}{3}$, are due to Kaplan, Lewenstein, Shafrir, and Sviridenko [JACM2005] and Elbassioni, Paluch, and van Zuylen [STACS 2012]. Using a result by Mucha [SODA 2013], which states that an $\alpha$-approximation algorithm for Max ATSP implies a $\bigl(2+\frac{11(1-\alpha)}{9-2\alpha}\bigr)$-approximation algorithm for the shortest superstring problem (SSP), we also obtain a $\bigl(2 \frac{33}{76} \approx 2.434\bigr)$-approximation algorithm for SSP. Furthermore, using an improved result by Englert et al. [STOC 2022], we achieve a $\approx 2.418$-approximation for SSP, beating the previous best-known approximation factor of $\approx 2.466$.

\end{abstract}

\newpage

\section{Introduction}
In the maximum asymmetric traveling salesman problem (Max ATSP),  also known informally as the ``taxicab ripoff problem'', we are given a complete directed graph $G=(V,E)$ (with  no loops or parallel edges) with nonnegative weights on the edges, and we wish to compute a traveling salesman tour (i.e., a Hamiltonian cycle) of maximum weight. Of course, the maximum asymmetric traveling salesman problem is equivalent to the minimum asymmetric traveling salesman problem (Min ATSP).  This is because a minimum-weight tour in a graph with a weight function $w: E \to \mathbb{R}_{\ge 0}$ corresponds to a maximum-weight tour in a graph with a weight function $w'$ defined so that for a suitably large constant $M$, every edge $e \in E$ satisfies $w'(e) = M - w(e)$.
However, in contrast to its minimization counterpart, Max ATSP can be approximated even when the edge weights do not satisfy the triangle inequality.
The current best approximation algorithms for it are due to Kaplan, Lewenstein, Shafrir,  Sviridenko \cite{KLSS} obtained in 2003 and Elbassioni, Paluch, van Zuylen \cite{PEZ} published in 2012.
Both of them achieve the approximation ratio of $\frac 23$, the former is based on linear programming and the other is  combinatorial and simpler. The problem is also known to be APX-hard \cite{PY}.

Max ATSP is interesting both from a theoretical point of view and because of its numerous applications. The most famous application is to the shortest superstring problem (SSP), which is defined as follows: we are given $n$ strings $s_1, s_2, \ldots, s_n$ over a given alphabet $\Sigma$, and we want to find a shortest string $s$ such that each $s_i$ is a substring of $s$ for all $1 \le i \le n$. SSP arises in DNA sequencing and data compression. Reductions from SSP to Max ATSP were shown, among others, by Mucha~\cite{Mucha}, who proved that an $\alpha$-approximation algorithm for Max ATSP implies a $\bigl(2+\frac{11(1-\alpha)}{9-2\alpha}\bigr)$-approximation algorithm for SSP. Improved reductions were recently shown by Englert, Matsakis, and Veselý in~\cite{EnglertStoc, EnglertIsaac23}.
Currently, the best approximation algorithm for SSP achieves an approximation ratio of $2.466$ via the reduction from~\cite{EnglertIsaac23}. For a long time, the best approximation algorithm for SSP was the one given by Sweedyk~\cite{sweedyk} in 1999, which achieved an approximation factor of $2 \frac{1}{2}$. The greedy conjecture for SSP states that a greedy algorithm—which in each step merges two strings with a maximum overlap (the longest suffix of one string that matches the prefix of another) over all pairs of strings and repeats this process until only one string remains—has an approximation factor of $2$. So far, the greedy algorithm has a proven approximation guarantee of $\approx 3.396$~\cite{EnglertIsaac23}.
Any $\alpha$-approximation algorithm for Max ATSP implies also an algorithm with the same guarantee for the maximal compression problem defined by Tarhio and Ukkonen \cite{TU}.

 We devise a  combinatorial $\frac{7}{10}$-approximation algorithm for Max ATSP, thus proving
\begin{theorem}
There exists a $\frac{7}{10}$-approximation algorithm for the maximum asymmetric traveling salesman problem.
\end{theorem}
Using the results in \cite{EnglertStoc}, \cite{EnglertIsaac23}, we obtain
\begin{corollary}
 
There exists a $\approx 2.4188$-approximation algorithm for the shortest superstring problem.
\end{corollary}

The presented results are a simpler and weaker version of \cite{34max}. A fully detailed version of \cite{34max} would be significantly longer
than the present paper. \footnote{To achieve a $\frac{3}{4}$-approximation one needs to compute two additional cycle covers $C_1, C_2$ with half-edges, because it is more difficult to get rid of tricky $2$-triangles (they cannot be replaced in the manner shown in this version). Additionally, it is necessary to get rid of $4$-cycles, whose two
subcycles belong to $C_{max}$.}

\vspace{0.5cm}
The approach we have adopted is as follows. We start by computing a maximum-weight \emph{cycle cover} $C_{\max}$ of $G$, where a cycle cover $C$ of a graph $G$ is defined as a set of directed cycles of $G$ such that each vertex of $G$ belongs to at most one cycle of $C$. (We allow vertices to be isolated.) A maximum-weight cycle cover of $G$ can be found in polynomial time via a reduction to the maximum-weight matching problem. 
Let $\mathit{opt}$ denote the weight of a maximum-weight traveling salesman tour of $G$. The weight of an edge $e$ will be denoted by $w(e)$, and for any subset of edges $E' \subseteq E$, its weight $w(E')$ is defined as $\sum_{e \in E'} w(e)$. Since a traveling salesman tour is a cycle cover of $G$ consisting of exactly one cycle, it follows that $w(C_{\max}) \ge \mathit{opt}$. 
By removing the lightest edge from each cycle of $C_{\max}$, we obtain a collection of vertex-disjoint paths, which can be arbitrarily patched to form a tour. Removing the lightest edge from a cycle $c$ of length $k$ results in a path of weight at least $\frac{k-1}{k} w(c)$. Because $C_{\max}$ may contain cycles of length two ($2$-cycles), in the worst case, the obtained tour may have a weight of only $\frac{1}{2} w(C_{\max})$. If we could compute a maximum-weight cycle cover of $G$ containing no $2$-cycles or $3$-cycles (also referred to as \emph{triangles}), we would achieve a $\frac{3}{4}$-approximation. Unfortunately, even finding a maximum-weight cycle cover without $2$-cycles is APX-hard~\cite{BM}.

\vspace{0.5cm}
{\bf Eliminating problematic  subgraphs with the aid of half-edges\ } 
Since $2$- and $3$-cycles in a maximum weight cycle cover are an obstacle to getting an approximation with ratio better than $\frac{2}{3}$, we would like to  somehow get rid of them. To this end we use a technique of eliminating problematic subgraphs with the aid of {\bf \em half-edges} - a half-edge of edge $(u,v)$ is informally speaking
``either a head or a tail of $(u,v)$''.  Half-edges were introduced in \cite{PEZ} and have been employed in \cite{12tsp}, \cite{maxtsp}, and \cite{01tsp}, among others. Here, we further develop this approach and show how to eliminate even more complex subgraphs. 

   We already know that computing a maximum weight cycle cover without $2$- and $3$-cycles is hard. What we propose instead is to find a cycle cover $C_1$ {\em improving on $C_{max}$} in the sense it does not contain certain $2$- and $3$-cycles from $C_{max}$ as well  as some other difficult subgraphs, but possibly contains half-edges and has weight at least $opt$. Such a cycle cover $C_1$ is  called a
{\bf \em relaxed cycle cover improving} $\bm{C_{max}}$. Let us note that it is the  requirement  that the weight of $C_1$ is an upper bound on $opt$ that makes the task difficult.  Without it, finding new cycle covers avoiding prescribed configurations is easy, and we would not even have to resort to half-edges. 
 We believe that the method utilizing half-edges provides a handy and relatively easy way
of obtaining new cycle covers (or sometimes matchings) improving on previous ones in a certain manner 
and having weights upper- or lower-bounding $\mathit{opt}$, respectively. Additionally, half-edges in such cycle covers can be either completely discarded or extended to full edges, yielding sets of directed paths and cycles with good properties. Such an approach is often substantially easier than extracting a good cycle cover from the fractional solution of an appropriate linear program. For example, note that the method of obtaining two cycle covers of weight at least $2\mathit{opt}$ and without any common $2$-cycle in \cite{KLSS} is very complicated.

\vspace{0.5cm}
{\bf Path-coloring \ } Let $\alpha$ denote a desired approximation factor. From $a_{\alpha}$ copies of $C_{max}$ and $b_{\alpha}$ copies of $C_1$ we obtain a multigraph $G_1$ with a total edge weight   at least 
$(a_{\alpha}+b_{\alpha})opt$.  Each occurrence of an edge $e$ in $C_{max}$ contributes $a_{\alpha}$ copies of $e$ to $G_1$, and each occurrence of $e$ in $C_1$ contributes $b_{\alpha}$ copies of $e$ to $G_1$. If $C_1$ contains only one half-edge of a certain edge $e$, then $C_1$ contributes $\frac{{b_\alpha}}{2}$ copies of $e$ to $G_1$.  
To extract a tour from $G_1$ with a weight of at least $\alpha \mathit{opt}$, we can choose a positive integer $k_{\alpha}$ such that $\alpha=\frac{a_{\alpha}+b_{\alpha}}{k_\alpha}$ and then \emph{\textbf{path-color}} $G_1$ with $k_{\alpha}$ colors. By this, we mean coloring each edge of $G_1$ with one of $k_{\alpha}$ colors so that edges of the same color form a collection of vertex-disjoint paths. (Thus, multiple copies of the same edge must be assigned distinct colors.) Next, we choose a color representing a set of edges with maximum weight, which is equal to at least $\frac{w(G_1)}{k_\alpha}$, where $w(G_1)$ denotes the total weight of edges in $G_1$. The paths from this set can be connected with the aid of other edges of $G$ to form a tour. Since the added edges  have nonnegative weights,  the obtained tour has a weight of at least $\frac{w(G_1)}{k_\alpha} \ge \alpha \mathit{opt}$. 
For example, to achieve a $\frac{3}{4}$-approximation, we might construct $G_1$ from one copy of $C_{\max}$ and two copies of $C_1$ and 
and then path-$4$-color it. 

In certain cases, such a multigraph $G_1$ - consisting of one copy of $C_{\max}$ and two copies of $C_1$ - would suffice to build a traveling salesman tour of weight at least $\frac{3}{4}\mathit{opt}$.
However, we can notice that we are unable to path-$4$-color it if $C_1$ contains a \emph{\textbf{$2$-triangle}} $t$, which is a triangle that shares an edge with a $2$-cycle of $C_{\max}$. This is due to the following.  Suppose that a triangle $t$ of $C_1$ contains the edges $(p,q), (q,r), (r,p)$ and a $2$-cycle of $C_{\max}$ contains the edges $(r,q), (q,r)$ (see Figure~\ref{2triangle}(a)). Then a subgraph of $G_1$ induced on the vertices of $t$ contains $2\cdot 3 + 1\cdot 2 = 8$ edges, one of which belongs to an edge oppositely oriented to an edge of $t$, namely $(r,q)$. A color assigned to $(r,q)$ cannot be used on the remaining edges. We need to use $3$ colors on $(q,r)$. A color assigned to $(q,r)$ may also be assigned to $(r,p)$ or $(p,q)$, but not both. We use $2$ of these colors for $(p,q)$ and the other one for $(r,p)$. This means that one copy of $(r,p)$ is left uncolored---we would need $5$ colors to path-color the whole subgraph of $G_1$ induced on $\{p,q,r\}$. Another obstacle arises when $C_1$ contains a $2$-cycle that shares one edge with $C_{\max}$. Note that, by definition, a $2$-cycle of $C_1$ cannot share both edges with $C_{\max}$. However, we can deal with such $2$-cycles by applying alternating paths from the symmetric difference of $C_1$ and $C_{\max}$.


To be able to handle $2$-triangles, we weaken the aimed-for approximation factor. We can deduce some general properties regarding  the number of copies $a_\alpha, b_\alpha$ that facilitate path-$k_{\alpha}$-coloring. Suppose that $a_\alpha \le b_\alpha$, $C_1$ contains an edge $e=(u,v)$, and $C_{\max}$ contains edges $e_1=(u,u'), e_2=(v',v)$. Then no color assigned to any copy of $e_1$ or $e_2$ may be assigned to any copy of $e$. If the copies of $e_1$ and $e_2$ are colored with disjoint sets of colors (yielding $2 a_\alpha$ different colors), we should have enough colors left for coloring the copies of $e$. Hence, we get a requirement that $2 a_\alpha + b_\alpha \le k_\alpha$. 
Since we allow half-edges in $C_1$, the number $b_\alpha$ should be even. Furthermore, $C_1$ may simultaneously contain  a half-edge of an edge $(u,v)$ incident to $u$ (such a half-edge is ``hanging over '' $v$) and a full edge $(w,v)$. Thus, in the multigraph $G_1$, we will have $\frac{b_{\alpha}}{2}$ copies of $(u,v)$, $b_{\alpha}$ copies of $(w,v)$, and $a_{\alpha}$ copies of some edge $(w',v)$ belonging to $C_{\max}$. Because all these edges must be colored differently, we require that $a_{\alpha} + \frac{3}{2}b_{\alpha} \le k_{\alpha}$. Additionally, we must be able to color a $2$-cycle of $C_1$, which implies that $2 b_{\alpha} \le k_{\alpha}$.
Consequently, choosing $\alpha=\frac{5}{7}$ with $a_\alpha=1$, $b_{\alpha}=4$, and $k_\alpha=7$ is unsuitable because coloring a $2$-cycle would require $2 \cdot 4=8>7$ colors. Moreover, for each vertex in the multigraph, we would like to allow for some flexibility rather than being forced to use all available colors for the incoming or outgoing edges. For these and other technical reasons, we opt for $\alpha=\frac{7}{10}$ with $a_\alpha=4$, $b_{\alpha}=10$, and $k_\alpha=20$. To obtain a path-$20$-coloring of $G_1$, we first (almost) path-$4$-color a multigraph $G^4_1$ consisting of $1$ copy of $C_{\max}$ and $2$ copies of $C_1$. Modifying this path-coloring of $G^4_1$ into a path-$20$-coloring of $G_1$ is then relatively straightforward.

We observe that a subgraph of $G_1$ induced by a $2$-triangle remains non-path-$20$-colorable. (The reason is similar as with path-$4$-coloring, more details are given  in Section~\ref{outline}.)
To safeguard against   $2$-triangles in $C_1$, we introduce a novel technique of \emph{\textbf{diluting}}. 
It consists in allowing a tricky triangle $t$ to occur in $C_1$, but in a {\em diluted} form, by which we mean that although it contains all
its edges, its weight is seemingly  decreased, which enables its path-coloring.

We are convinced that the presented techniques will find many other applications beyond the traveling salesman problem. They can be helpful in graph problems where specific subgraphs must be forbidden. For example, they have already been used in the triangle-free $2$-matching problem in non-bipartite graphs and the square-free $2$-matching problem in bipartite graphs.

\vspace{0.5cm}
{\bf Method of path-coloring \ } 
For coloring $G_1$, we present a method that we think is interesting in its own right. One of the surprisingly simple ideas on which this method is based is as follows: let $S$ be a subset of $V$ and $e=(u,v)$ an edge going into $S$ (i.e., $u \notin S$ and $v \in S$) colored with a color $k$. If there exists no edge $e'=(u',v')$ outgoing from $S$ (i.e., such that $u' \in S$ and $v' \notin S$) that is colored $k$, then $e$ does not belong to any cycle whose edges are all colored $k$. We use this idea inductively to path-color the multigraph $G_1$.  

The multigraph coloring considered in this paper is also related to \emph{the linear arboricity conjecture}, which asserts that every $k$-regular digraph can be path-$(k+1)$-colored \cite{alon, Naka}. Our methods of path-coloring provide simple, easy-to-prove algorithms for path-$3$-coloring a $2$-regular digraph and path-$4$-coloring a $3$-regular digraph consisting of $2$ copies of one cycle cover and $1$ copy of another. (These algorithms are concise special cases of the general path-coloring algorithm presented in this paper. For comparison, the algorithm for path-$3$-coloring a $2$-regular digraph shown in \cite{KLSS} is quite complicated and has a long analysis.)

\vspace{0.5cm}
{\bf Previous and related results}
The history of approximating the problems of maximum asymmetric traveling salesman and shortest superstring is quite long as  is shown by the following lists of papers  \cite{FNW}, \cite{KPS} \cite{B1}, \cite{LS}, \cite{KLSS},
\cite{PEZ} and 
 \cite{Li},  \cite{Blum},  \cite{Teng}, \cite{Czumaj},  \cite{KPS}, \cite{armen95}, \cite{armen95}, \cite{BJJ}, \cite{sweedyk}, \cite{KLSS}, \cite{PEZ}, \cite{Mucha},
\cite{EnglertStoc}, \cite{EnglertIsaac23}.

Other variants of the maximum traveling salesman problem that have been considered are among others: the maximum symmetric traveling salesman problem (MAX TSP), in which the underlying graph is undirected - currently the best known approximation ratio is $\frac 45$ \cite{maxtsp}, the maximum  metric symmetric traveling salesman problem - the best approximation factor is $\frac 78$ \cite{KM2},
the maximum asymmetric traveling salesman problem with a triangle inequality - the best approximation ratio is $\frac{35}{44}$ \cite{KM1}.
 
\section{Outline of algorithm} \label{outline}
Suppose we have computed a maximum-weight cycle cover $C_{\max}$ of a given complete directed graph $G=(V,E)$. We say that a cycle $c$ is \emph{\textbf{$\alpha$-hard}} if it belongs to $C_{\max}$ and each edge $e \in c$ satisfies $w(e) > (1-\alpha)w(c)$. We refer to cycles of length $i$ (i.e., consisting of $i$ edges) as \emph{\textbf{$i$-cycles}}; in particular, $3$-cycles are called \emph{\textbf{triangles}}. 
Notice that for $\alpha$ satisfying $\frac{1}{2} < \alpha \le \frac{3}{4}$, only $2$-cycles and triangles can be $\alpha$-hard. By $c=(v_1, v_2, \dots, v_i)$ we denote an $i$-cycle consisting of the edges $(v_1, v_2), \dots, (v_{i-1},v_i), (v_i, v_1)$. If $C_{\max}$ does not contain any $\alpha$-hard cycles, then we can easily construct a traveling salesman tour of weight at least $\alpha \cdot w(C_{\max}) \ge \alpha \cdot \mathit{opt}$. 
Conversely, if $C_{\max}$ contains at least one $\alpha$-hard cycle, we would like to find an alternative cycle cover $C_1$ that contains no $\alpha$-hard cycles from $C_{\max}$ (meaning that for each $\alpha$-hard cycle $c \in C_{\max}$, at least one edge of $c$ is absent from $C_1$). Furthermore, $C_1$ should have a total weight of at least $\mathit{opt}$ and exhibit  properties that facilitate finding a tour of weight at least $\alpha \cdot \mathit{opt}$. Let us note that computing a cycle cover of weight at least $\mathit{opt}$ that completely excludes all $\alpha$-hard cycles is rather infeasible; for comparison, finding a maximum-weight cycle cover without any $2$-cycles is known to be NP-hard~\cite{BM}. To circumvent this difficulty, we relax the definition of a cycle cover to permit the inclusion of \emph{\textbf{half-edges}}.  A half-edge of edge $(u,v)$ is informally speaking
``half of the edge $(u,v)$ that contains either a head or a tail of $(u,v)$''. We formally define half-edges and cycle covers allowing half-edges later. For now one may think of $C_1$ as a standard cycle cover.

After admitting half-edges, the computation of a maximum weight cycle cover without some forbidden cycles becomes possible. 
This is roughly done by preventing one edge on each forbidden cycle from appearing in a cycle cover. (We observe that using this approach we cannot get a cycle cover without any triangle, because then we would have to block too many edges.) We forbid  some triangles and $2$-cycles of $C_{max}$ and use both $C_{max}$ and $C_1$ to construct a solution. Suppose that $\alpha=\frac{a_{\alpha}+b_{\alpha} }{k_{\alpha}}$ and  $a_{\alpha}, b_{\alpha}, k_{\alpha}$  are positive integers. To extract  a tour of weight at least $\alpha \cdot opt$ from $C_{max}$ and $C_1$, we are going to build a multigraph $G_1$ consisting of $a_{\alpha}$ copies of $C_{max}$ and $b_{\alpha}$ copies of $C_1$. More precisely, $G_1$ contains $a_{\alpha}$ copies of  each edge $e \in C_{max}\setminus C_1$, $b_{\alpha}$ copies of each   $e \in C_1\setminus C_{max}$ and $a_{\alpha}+b_{\alpha}$ copies of each $e \in C_1 \cap C_{max}$.  We would like to color each edge of $G_1$ with one of $k_{\alpha}$ colors   so that  edges of the same color form a collection of disjoint paths or, in other words, we would like to \emph{\textbf{path-$k_{\alpha}$-color}} $G_1$ (or \emph{\textbf{path-color it with $k_{\alpha}$ colors}}).
By $G^{20}_1$ and $G^4_1$ we denote a  multigraph $G_1$ created in the above manner, respectively, (i) for $\alpha=\frac{7}{10}$ and $a_{\alpha}=4, \ b_{\alpha}=10, k_{\alpha}=20$ and (ii) for  $\alpha=\frac{3}{4}$ and $a_{\alpha}=1, \ b_{\alpha}=2, k_{\alpha}=4$.  To obtain a path-coloring of  $G^{20}_1$ (with $20$ colors), we first path-color with $4$-colors a multigraph $G^4_1$.

\begin{figure}[h]
\centering{\includegraphics[scale=0.9]{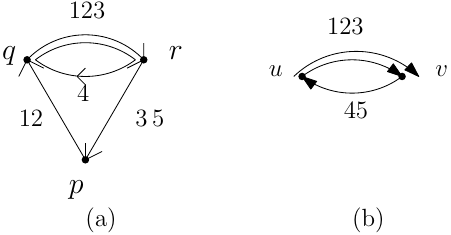}}
\caption{ (a) A triangle $(p,q,r)$ sharing an edge with a $2$-cycle $(q,r)$, denoted in short as $(p;q,r)$. (b) A $2$-cycle $(u,v)$ sharing one edge with a cycle of $C_{max}$. }
 \label{2triangle}
\end{figure}

A cycle of $G$ oppositely oriented to $c$ is denoted as $opp(c)$.   A cycle $c'$ is said to be a {\bf \em subcycle} of $c$ if every vertex of $c'$ belongs to $c$.
We may notice that path-$4$-coloring of $G^4_1$  is not possible, if
$C_1$ contains one of the following:
\begin{enumerate}
\item a $2$-cycle that either belongs to $C_{max}$ or is a subcycle of a triangle of $C_{max}$.
\item a  triangle $t$, called a {\bf \em   $\mathbf \mathit 3$-triangle} such that either $t$ or   $opp(t)$ belongs  to $C_{max}$. 

\item a triangle $t=(p,q,r)$ such that a $2$-cycle $(q,r)$ belongs to $C_{max}$, denoted $(p;q,r)$ and called  a {\bf \em   $\mathbf \mathit2$-triangle}. See Figure \ref{2triangle} (a). Let us note that a $2$-triangle $(p;q,r)$ is non-path-$20$-colorable in $G^{20}_1$ either. The reason is similar as in $G^4_1$. We need to use $14$ colors on $(q,r)$. A color assigned to $(q,r)$ may be also assigned to $(r,p)$ or $(p,q)$ but not both. We use $7$ of these colors for $(p,q)$ and $7$ for $(r,p)$. We assign each of $15,16,17$ to both $(r,p)$ and $(p,q)$. This means that we are left with only $3$ colors for $(r,q)$ but we need $4$.

\item a $2$-cycle $c=(u,v)$ such that one of its edges belongs to $C_{max}$. This is because $G^4_1$ contains in this case $5$ edges connecting
$u$ and $v$ ($3$ in one direction and $2$ in the other). See Figure \ref{2triangle} (b).

\end{enumerate}

We later show that if $C_1$ does not contain any of the above cycles, then $G^{20}_1$ is always path-$20$-colorable and $G^4_1$ almost path-$4$-colorable.  Ideally, we would like the enumerated cycles not to
occur in $C_1$ at all.  However, not all of them are bad for our purposes, because sometimes it is easy to replace some edges of these cycles
with other ones, so that we obtain a path-$4$-colorable multigraph. For example, if $C_1$ contains a $2$-triangle $t=(p; q,r)$ and $w(r,q) \leq w(q,r)$, we can replace $1$ copy of $(r,q)$ with $1$ copy of $(q,r)$ and 
make the subgraph on $p,q,r$  path-$4$-colorable.
Also, this way  we do not diminish the overall weight of the subgraph.

In Section~\ref{trickysec}, we formally define a set of cycles that are \emph{\textbf{tricky}}. Informally, a cycle $c$ is tricky if it is of type $1$, $2$, or $3$ described above, and its occurrence in $C_1$ implies that $G_1$ is not path-colorable and this cannot be remedied via local edge replacements. (The set of permitted local replacements is finite and relatively small.) As for $2$-cycles of type $4$ (which are not also of type $1$), we prove in Section~\ref{completing} that the subgraphs induced on them can always be made path-colorable by applying a local edge exchange.
 For any multisubgraph $G'$ of $G$, by $mult_{G'}(e)$ we denote the number of copies of $e$ occurring in $G'$.

We observe  that tricky $3$-triangles are vertex-disjoint.  Similarly, a tricky $2$-cycle of $C_{max}$ is vertex-disjoint with any other tricky $2$-cycle or $3$-triangle. On the other hand tricky $2$-triangles may overlap - if $c$ is a $2$-cycle of $C_{max}$, then there may be many  tricky $2$-triangles sharing the same edge of $c$. Thus, we cannot forbid all tricky $2$-triangles in $C_1$, because similarly as with the set of all triangles we would have to block too many edges. Instead, from the set of all tricky $2$-triangles we choose a  subset $R$ of their representatives and forbid only them. The set of {\bf \em problematic} cycles of $G$ consists of all  tricky $2$-cycles,  tricky $3$-triangles,  and  tricky $2$-triangles of $R$. These are cycles that will not appear in $C_1$.

To be able to compute a cycle cover $C_1$ improving $C_{max}$,
we allow it to contain {\em half-edges}, defined as follows. Let $\tilde G =(\tilde V, \tilde E)$  be a graph obtained from $G$ by splitting 
 each edge  $(u,v)\in E$ with a vertex $x_{(u,v)}$  into two edges $(u,x_{(u,v)})$ and $(x_{(u,v)},v)$ having weights such that $w(u,x_{(u,v)})+ w(x_{(u,v)},v)=w(u,v)$. (The distribution of weights is given later.) Each of the edges $(u, x_{(u,v)}),  (x_{(u,v)},v)$ is called
{\bf \em a half-edge (of $(u,v)$)}.  By  saying that an edge $(u,v)$ of $G$ belongs to a subset  $\tilde C \subseteq \tilde E$, we  mean that both half-edges of $(u,v)$ belong to $\tilde{C}$. We  say that $\tilde C \subseteq \tilde E$ does not contain a  cycle $c$ of $G$, if $\tilde C$ does not contain all edges  of $c$, i.e., there exists at least one edge $e$ of $c$  that  does not belong to $\tilde C$.

We first define a quasi relaxed cycle cover   improving $C_{max}$. Such a cycle cover may contain tricky $2$-triangles of $R$.

\begin{definition}\label{rel2prim}
A {\bf \em quasi relaxed cycle cover   improving $C_{max}$}  is a subset $\tilde C\subseteq \tilde E$ such that
\begin{itemize}
\item[(i)]
each vertex in $V$ has exactly one outgoing and one incoming half-edge in $\tilde C$;
\item[(ii)] for any problematic cycle $c$, which is not a tricky $2$-triangle of $R$, $\tilde C$  does not contain   $c$; 
\item[(iii)] if $\tilde C$ contains only one half-edge of edge $(u,v)$, then $(u,v)$ belongs to a problematic cycle, which is not a tricky $2$-triangle of $R$. 
\end{itemize}
\end{definition}

\vspace{0.5cm}
A quasi relaxed cycle cover $C$  improving $C_{max}$ consists of directed cycles and/or  directed paths. A directed cycle of   $C$ corresponds to a directed cycle of the original graph $G$
and a directed path ends and begins with a vertex in $\tilde V \setminus V$. \\


\subsection{Tricky $2$-triangles}
We now describe how tricky $2$-triangles are handled. Let $t = (p, q, r)$ be a tricky $2$-triangle. We call $p$ its \emph{\textbf{t-point}} and the $2$-cycle $(q,r)$ its \emph{\textbf{t-cycle}}. To the $2$-cycle $c$, we assign a modified weight $w'(c) = w(q,r)$ 
The motivation behind this weight $w'$ is that, as we later show, a greater weight $w'$ of the t-cycle $(q,r)$ of a tricky $2$-triangle $t$ implies a greater weight of the edges of $t$. It is precisely with the edges of triangles in $R$ that we replace some of the edges of tricky $2$-triangles occurring in $C_1$ within $G_1$, thereby making these subgraphs of $G_1$ path-$20$-colorable.



To identify the set $R$, we proceed as follows. We construct a bipartite graph $H = (C \cup P, E_t)$, where the vertex set $C$ represents the t-cycles and $P$ represents the t-points. An edge $(c,p)$ belongs to $E_t$ if and only if there exists a tricky triangle $t$ such that $c$ is its t-cycle and $p$ is its t-point.  Let $T_1 \cup T_2 \cup \dots \cup T_k$ be a partition of the vertex set $C$ such that:
\begin{enumerate}
    \item[(i)] for each $i$, any two t-cycles of $T_i$ have equal weight $w'$, and
    \item[(ii)] for any $i, j$ such that $i < j$, and any $c_i \in T_i, c_j \in T_j$, it holds that $w'(c_i) > w'(c_j)$.
\end{enumerate}
We assign \emph{ranks} to the edges of $H$ in the following manner: any edge in $E_t$ incident to a vertex in $T_i$ has rank $i$.
We then compute a \emph{\textbf{rank-maximal}} matching $N$ of $H$, which is a matching that contains the maximum possible number of rank-one edges, and subject to this condition, the maximum number of rank-two edges, and so on. A rank-maximal matching can be computed in polynomial time~\cite{rank}.

We can  observe that 
\begin{fact}
Any rank-maximal matching of $H$ is a maximum matching of $H$.
\end{fact}
\dowod 
Briefly, it follows from the fact that all edges incident to a given vertex in $C$ have the same rank. In more detail, suppose for contradiction that there exists a rank-maximal matching $M$ of $H$ that is not a maximum matching. Hence, $H$ contains an $M$-augmenting path $P = (c_1, p_1, \dots, c_k, p_k)$. However, the matching $M' = M \oplus P$ then satisfies the property that for each rank $i$, the number of edges of rank $i$ in $M'$ is greater than or equal to the number of edges of rank $i$ in $M$. Because $M'$ contains strictly more edges than $M$, this implies that $M$ is not rank-maximal - a contradiction.
\koniec


As the set $R$ representing tricky $2$-triangles we set tricky triangles corresponding to the edges of the rank-maximal matching $N$. In $C_1$, each triangle of $R$ is either eliminated (i.e., not contained) or diluted (as defined below), whereas any other tricky $2$-triangle may appear unaltered. If any tricky $2$-triangles do occur in $C_1$, we replace some of their edges in $G_1$ with the edges of overlapping triangles from $R$.

To ensure that no triangle of $R$ appears in $C_1$, one might consider a similar approach to the one used for tricky $3$-triangles, requiring that at least two half-edges of any triangle in $R$ do not occur in $C_1$. It turns out, however, that this condition is insufficient because we may still obtain subgraphs on $t$ that are not path-colorable. 
To tackle the triangles in $R$, we introduce a technique stronger than the one with half-edges; namely, we propose to \emph{\textbf{dilute}} them.  Specifically, for a triangle $t=(p;q,r) \in R$, let $\kappa(t)$ denote a parameter specified later. We allow $t$ to appear in $C_1$ only if its weight is \emph{seemingly} diminished by $\kappa(t)$—that is, only if the weight of $t$ in $C_1$ equals $w(t)-\kappa(t)$. At first glance, this seems impossible because the edge weights in the original graph $G$ are fixed. Nevertheless, such a trick can be realized by conducting the computation of a relaxed cycle cover (improving upon $C_{\max}$) in a modified graph $G'$, which is in a way a supergraph of $\hat{G}$, defined below. The graph $G'$ contains a so-called ``gadget'' for each problematic cycle. The ability to obtain a diluted triangle $t$ then arises from reducing the weight of certain edges of $t$ and transferring this weight to auxiliary edges within the corresponding gadget. More details are provided in the following section.

Diluting triangles is helpful because we choose $\kappa(t)$ in such a way that it effectively reduces the number of copies of the edge $(r,q)$ in the multigraph $G_1$ by one compared to if $t$ were not diluted, which enables the subgraph on $t$ to be path-colored.

To encode the idea of diluting triangles of $R$ in the context of the graph $\tilde{G}$, we extend $\tilde{G}$ as follows. For each tricky triangle $t \in R$, we add two new vertices $v_t$ and $v'_t$, along with two loops: $e_t$ incident to $v_t$ and $e'_t$ incident to $v'_t$, with weights $w(e_t) = -\kappa(t)$ and $w(e'_t) = \kappa(t)$, respectively. We denote this resulting graph by $\hat{G} = (\hat{V}, \hat{E})$. Note that this is a supergraph of $\tilde{G}$. For a triangle $t = (p,q,r) \in R$, we say that $t$ is \emph{\textbf{diluted}} in $C_1$ if $C_1$ contains both $t$ and the loop $e_t$, but does not contain $e'_t$. This implies that the total weight of $t$ in $C_1$ can be interpreted as being equal to $w(t) - \kappa(t)$. We are now ready to provide a complete definition of a relaxed cycle cover that improves upon $C_{\max}$.

\begin{definition}\label{rel2}
A {\bf \em relaxed cycle cover   improving $C_{max}$}  is a subset $\hat C\subseteq \hat E$ such that
\begin{itemize}
\item[(i)]
each vertex in $V$ has exactly one outgoing and one incoming half-edge in $\hat C$;
\item[(ii)] any problematic cycle $c$ contained in $\hat C$ is a diluted triangle of $R$;
\item[(iii)] if $\hat C$ contains only one half-edge of edge $(u,v)$, then $(u,v)$ belongs to a problematic cycle.

\end{itemize}
\end{definition}

The outline of a $\frac{7}{10}$-approximation algorithm for Max ATSP is as follows.

\begin{algorithm}[H]
  \begin{algorithmic}[1]
    
    \State \label{} Compute a maximum weight cycle cover $C_{max}$ of $G$.
    \State \label{} If $C_{max}$ does not contain any hard cycle, extract from $C_{max}$ a set ${\cal P}$ of vertex-disjoint paths of weight at least $\frac{7}{10}w(C_{max})$ and go to  \ref{itm:step6}.
    \State \label{} Compute a relaxed cycle cover $C_1$ improving $C_{max}$ with weight $w(C_1) \geq opt$.
    \State \label{} Compute a multigraph $G^{20}_1$ with weight $w(G^{20}_1) \geq 4w(C_{max})+10 w(C_1)$ and path-$20$-color $G^{20}_1$ omitting non-path-$20$-colorable subgraphs. If the whole $G^{20}_1$ is path-colored, go to  \ref{itm:step6}.
		\State Compute subsets $E_1, F_1 \subset E$, called {\em exchange sets}, such that $G_2=G^{20}_1 \setminus E_1 \cup F_1$ is path-$20$-colorable. Extend the existing  coloring of $G^{20}_1$ to that of $G_2$.
    \State \label{itm:step6} Extend a set ${\cal P}$ of vertex-disjoint paths of weight at least $\frac{7}{10}opt$ to a tour $\mathcal T$ of $G$. \\
    \Return $\mathcal T$
  \end{algorithmic}
	\caption{A $7/10$-approximation algorithm for Max ATSP.}
\label{alg:76approx}
\end{algorithm}

\section{Computation of a relaxed cycle cover $C_1$}

\subsection{Main ideas}
To compute a quasi relaxed cycle cover  $C_1$  improving $C_{max}$  we construct the following undirected  graph $G'=(V',E')$.
For each vertex $v$ of $G$ we add two vertices $v_{in}, v_{out}$ to $V'$. For each edge $(u,v)$ that belongs to a problematic cycle,
we add vertices $e^1_{uv}, e^2_{uv}$, called {\bf \em subdivision vertices} of $(u,v)$, an edge  $(e^1_{uv}, e^2_{uv})$ of weight $0$ and edges $(u_{out}, e^1_{uv}), (v_{in}, e^2_{uv})$ having weights such that $w(u_{out}, e^1_{uv}) +w(v_{in}, e^2_{uv})= w(u,v)$. Edges $(u_{out}, e^1_{uv}), (v_{in}, e^2_{uv})$ are also called {\bf \em half-edges} of $(u,v)$. For every other  edge $(u,v) \in E$   we add an edge $(u_{out}, v_{in})$ of weight $w(u,v)$.

Next we build so-called gadgets for problematic cycles. The main ideas behind them are the following. If a problematic cycle $c$ has length $|c|$, then the edges forming it contain $2|c|$ half-edges. In order for $c$ not to occur in $C_1$, we construct a gadget in which we prevent at least two half-edges of $c$ from appearing in $C_1$. It may happen that a gadget blocks one half-edge of edge $e_1 \in c$ and one half-edge of a different edge $e_2$ of $c$.  Sometimes, we may also want to forbid a cycle $opp(c)$ from occurring in $C_1$ and then we block  two more half-edges contained in the edges of $opp(c)$ in the gadget. From this collection of edges
and half-edges we later build a multigraph $G^{20}_1$. To make this multigraph easier to handle, we want the configurations of half-edges within gadgets to satisfy certain properties. A very important one of them is that an indegree and outdegree of each vertex is appropriately bounded. Another one used in gadgets for triangles is that
the weights of the half-edges forming an edge are distributed in a way that enables exchanges of half-edges (so-called flipping of half-edges.) 

\subsection{Technical description of gadgets}
\begin{figure}[h]
\centering{\includegraphics[scale=0.8]{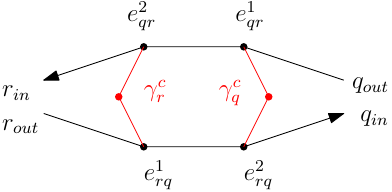}}
\caption{{\scriptsize  A gadget for a $2$-cycle $(q,r)$ .}
} \label{2cykl}
\end{figure}

For each tricky $2$-cycle $c$ on vertices $r$ and $q$,   we add vertices $\gamma^c_r$ and $\gamma^c_q$
and edges \\ $(\gamma^c_{r}, e^1_{rq}),  (\gamma^c_{r}, e^2_{qr}), (\gamma^c_{q}, e^1_{qr}), (\gamma^c_{q}, e^2_{rq})$ with weight $0$. 
The gadget is shown in Figure \ref{2cykl}. If $c$ is not a subcycle of a tricky $3$-triangle or $2$-triangle of $R$, each of the half-edges
of $(r,q)$ gets weight $\frac{1}{2}w(r,q)$ and each of the half-edges
of $(q,r)$ gets weight $\frac{1}{2}w(q,r)$.

Before describing the remaining gadgets, let us check what a perfect matching of $G'$ looks like on gadgets for tricky $2$-cycles and give a few definitions.  We  say that a half-edge $e_h$ is {\bf \em within a set of edges} $F\subseteq E$  if $e_h$ is a half-edge of some edge $e$ in $F$.
 Let $e_1, e_2$ denote two different edges of $G$ incident with the same vertex $v$. If $\tilde{C}$ contains only one half-edge of each of $e_1, e_2$, then these  half-edges 
are called {\bf \em crossing} if exactly one of them is incident to $v$ and
{\bf \em non-crossing} otherwise. A half-edge of $(u,v)$ is {\bf \em incoming} if it is  incident to $v$ and  {\bf \em outgoing} if it is incident to $u$. In Lemma \ref{relax} we prove that a  perfect matching of $G'$ yields a relaxed cycle cover $\tilde C$ that satisfies condition (1) for each tricky $2$-cycle. The proof is the same as in \cite{PEZ} and follows from the fact that in the gadget for a tricky $2$-cycle $c=
(r,q)$ the vertices $\gamma^c_r$ and $\gamma^c_q$ are either (i) matched to the subdivision vertices of the same edge of $c$, say $(r,q)$ (and then this edge $(r,q)$ is excluded from the relaxed cycle cover $\tilde C$) and the subdivision vertices of $(q,r)$ are either matched to each other (then none of the edges of $c$ belongs to $\tilde C$) or are matched to $q_{out}, r_{in}$ (which means that $(q,r) \in \tilde C$)
or (ii) are matched to subdivision vertices belonging to different edges of $c$ and then the remaining two subdivision vertices are matched either to $r_{out}$ and $q_{out}$ or to $r_{in}$ and $q_{in}$.

\begin{figure}
\begin{subfigure}[b]{0.55\textwidth}
\includegraphics[scale=0.9]{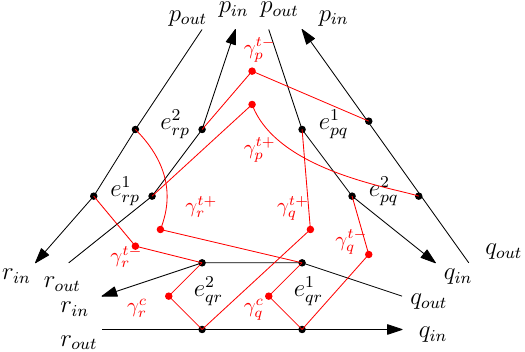}
\caption{{\scriptsize  A gadget for a $3$-triangle $t= (p,q,r)$.}
} \label{gtriangle3}

\end{subfigure} 
\begin{subfigure}[b]{0.4\textwidth}
\raisebox{0.5cm}{\includegraphics[scale=0.9]{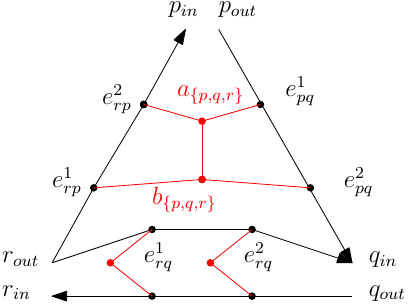}}
\caption{{\scriptsize  A gadget for a $2$-triangle $t= (p,q,r)$.}
} \label{gtriangle2}

\end{subfigure}
\caption{}

\label{gtriangle}
\end{figure}

\begin{figure}
\centering{\includegraphics[scale=0.6]{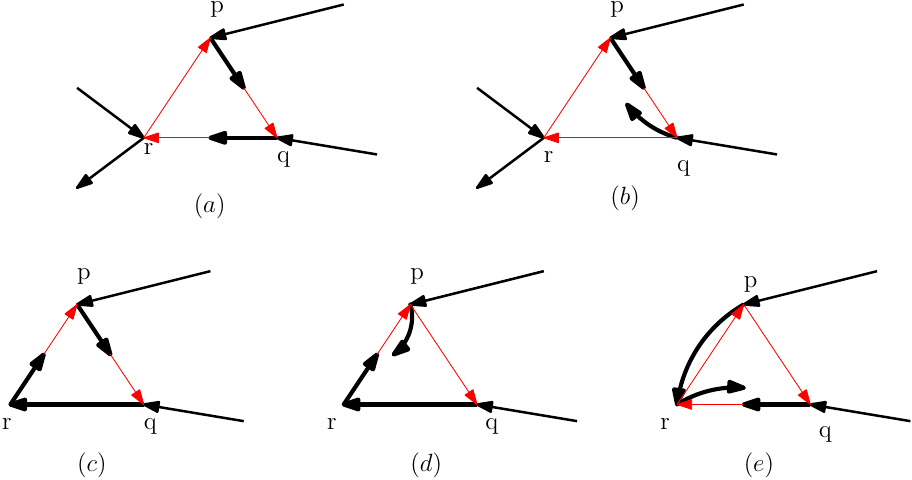}}
\caption{{\scriptsize  Configurations of half-edges within a tricky $3$-triangle $t$. In cases (b),(d), (e) we replace in $C_{max}$ the triangle $t$
with a $2$-cycle, correspondingly: $(p,q)$, $(p,r)$ and $(r,q)$.}
} \label{rozklad}
\end{figure}

Next we describe gadgets for tricky $3$-triangles.
Let $t$ be any tricky $3$-triangle $t=(p,q,r)$.  Among edges of $opp(t)$ we choose  one with maximum weight. Suppose that it  is $(r,q)$.  For each such  $t$, we proceed as follows.
We add vertices $\gamma^{t-}_p, \gamma^{t+}_p$ and connect them  to vertices $e^2_{qp}, e^2_{rp}$ and  $e^1_{qp}, e^1_{rp}$, respectively, 
via edges of weight $0$. Each of the edges  $(e^2_{qp}, p_{in}), (e^2_{rp}, p_{in})$ gets weight $\frac{1}{2}\max\{w(q,p), w(r,p)\}$. Thus,
$w(q_{out}, e^1_{qp})= w(q,p) - \frac{1}{2}\max\{w(q,p), w(r,p)\}, \ w(r_{out}, e^1_{rp})= w(r,p) - \frac{1}{2}\max\{w(q,p), w(r,p)\}$. 
We proceed analogously for pairs of edges $(r,q), (p,q)$ and $(p,r), (q,r)$.
Thus, we add vertices $\gamma^{t-}_q, \gamma^{t+}_q$ and connect them  to vertices $e^2_{pq}, e^2_{rq}$ and  $e^1_{pq}, e^1_{rq}$, respectively, 
via edges of weight $0$, and  we add vertices $\gamma^{t-}_r, \gamma^{t+}_r$ and connect them  to vertices $e^2_{pr}, e^2_{qr}$ and  $e^1_{pr}, e^1_{qr}$, respectively, 
via edges of weight $0$. Each of the edges  $(e^2_{pq}, q_{in}), (e^2_{rq}, q_{in})$ gets weight $\frac{1}{2}\max\{w(p,q), w(r,q)\}$ and each of the edges  $(e^2_{pr}, r_{in}), (e^2_{qr}, r_{in})$ gets weight $\frac{1}{2}\max\{w(p,r), w(q,r)\}$.
The $2$-cycle $c=(r,q)$ is  tricky, hence we have already added a gadget for it.
The gadget for $t$ is depicted in Figure  \ref{gtriangle3}. The unequal distribution of weights among half-edges and other elements of construction are to facilitate  the  rearrangement  of  half-edges
within  $t$ so that they  form one of  only a few manageable configurations shown in Figure \ref{rozklad} - the details are given in   Lemma \ref{relax}.
Also, to enable the flipping of half-edges, either all incoming half-edges for a given vertex in the gadget have the same weight or all outgoing half-edges have equal weight.

We say that a (quasi) relaxed cycle cover $\tilde C$ is {\bf \em non-integral} on a set $F$ of edges if there exists an edge $e \in F$ such that $\tilde C$ contains exactly one half-edge of  $e$. For a triangle $t$, the total weight of half-edges of $\tilde C$ within $t \cup opp(t)$ is denoted as $w(\tilde C)_t$.
An edge $(u,v)$ is said to be incoming to  (resp. outgoing of)  a cycle $c$ if $v \in c$ and $u \notin c$ (corr. $u \in c$ and $v \notin c$).

\begin{lemma}\label{relax}
Any perfect matching of $G'$ yields  a quasi relaxed cycle cover   $\tilde C$ with the following properties:

\begin{enumerate}
\item[(1)]
for each tricky $2$-cycle $c=(u,v)$, if $\tilde C$ contains two half-edges within $c$, then they either belong to the same edge or  are crossing (i.e., are either both incoming or both outgoing).  If $c$ is not a subcycle of a tricky triangle,  then $\tilde C$  contains either zero or    two half-edges within $c$.

\item [(2)] for each tricky  $3$-triangle $t=(p,q,r)$, if $\tilde C$  is non-integral on $t \cup opp(t)$, then: 
\begin{enumerate}
\item [(i)] the number of half-edges of $\tilde C$ within $t \cup opp(t)$ is even and at most $4$, 
\item[(ii)] the difference between the numbers of edges of $\tilde C$ incoming to $t$ and outgoing of $t$ is equal to exactly two 
and 
\item [(iii)] $10w(\tilde C)_t +4w(t)$ is upper bounded
by: 
\begin{itemize}
\item 
 $\max\{10w(t) -5w(r,p), 
10(w(p,q)+w(q,p))\}$, if $\tilde C$ contains one edge ougoing of $t$, incident to $r$ and three edges incoming to $t$, (this corresponds to 
a configuration in which the frayed half-edges are crossing within: $(p,q), (q,r)$ or  a $2$-cycle $(p,q)$)
\item $ \max\{10w(t) -5w(q,r), 
10(w(p,q)+w(q,p))\}$, if $\tilde C$ contains one edge incoming to $t$ incident to $r$ and three edges outgoing of $t$, 

\item $\max\{10w(t) +5w(q,r), 10( w(q,r)+w(r,p)+w(p,r)), 10( w(p,r)+w(r,q)+w(q,r))\}$, if $\tilde C$ contains two edges incoming to $t$ incident to $p$ and $q$ and no edge outgoing of $t$, 
\item 
$\max\{10w(t) +5w(r,p), 10( w(p,r)+w(p,q)+w(q,p)), 10(w(p,q)+w(p,r)+w(r,p))\} $, if $\tilde C$ contains two edges outgoing of $t$, incident to $p$ and $q$ and no edge incoming  to $t$.

\end{itemize}
\end{enumerate}

\end{enumerate}
\end{lemma}
The proof is in Section \ref{miss}. \\


To compute a relaxed cycle cover, i.e. one that satisfies  Definition \ref{rel2}, we  extend and modify $G'$, creating thus $G''$ as follows. For each tricky $2$-triangle $t=(p,q,r)$ of $R$ such that $(q,r)$ is its t-cycle, we add the following gadget. We add vertices $a_{\{p,q,r\}}, b_{\{p,q,r\}}$ and connect them  to vertices $e^1_{pq},  e^2_{rp}$ and $e^2_{pq}, e^1_{rp}$, respectively, 
via edges of weight $\frac{\kappa(t)}{2}$ and connect $a_{\{p,q,r\}}$ and $b_{\{p,q,r\}}$ via an edge of weight $0$. We also decrease the weight of each of the edges $(r_{out}, e^1_{rq}), (e^2_{rq}, q_{in}),  (q_{out}, e^1_{qr}), (e^2_{qr}, r_{in})$ by  $\frac{\kappa(t)}{2}$, where $\kappa(t)=\frac{w(r,q)}{10}$.


\begin{lemma}\label{tricky}

Any  perfect matching of $G''$ yields  a relaxed cycle cover  $\hat C$ with the following properties.
Let $t=(p,q,r)$ be a tricky $2$-triangle of $R$, where $c=(q,r)$ is its t-cycle. 

\begin{enumerate}
\item[(i)] If $\hat C$ is non-integral on $t \cup c$, then it contains
 (i) crossing half-edges  within $c$ or (ii) crossing half-edges within $\{(r,p), (p,q)\}$.

 \item[(ii)] If $\hat C$ contains  $t$, then  $t$  is diluted.
 \item[(iii)] If $\hat C$ contains a loop  $e'_t$, then it contains no half-edges within $c$. 
\end{enumerate}

\end{lemma}
The proof is in Section \ref{miss}.

\begin{theorem}
Assume that $G$ has at least $4$ vertices.
Any perfect matching of $G''$ yields  a relaxed cycle cover  $C_1$ improving $C_{max}$.
A maximum weight perfect matching of $G''$ yields a relaxed cycle cover $C_1$ improving $C_{max}$  such that $w(C_1) \geq opt$.
\end{theorem}
\dowod The first statement follows from the preceding two lemmas.
The second statement follows from the fact that  a traveling salesman tour is also a relaxed cycle cover improving $C_{max}$. \koniec

\section{Path-coloring} \label{pathcol}

Once we have computed a maximum weight cycle cover $C_{max}$ and a relaxed cycle cover $C_1$ our next task is to color a multigraph $G_1$. 
We say that a tricky cycle $c$ is {\bf \em halfy} if $C_1$ contains exactly one half-edge of some edge of $c$ or of $opp(c)$.
In this section we assume that $C_1$ may contain only halfy $2$-cycles of $C_{max}$.
Two edges $e_1, e_2$ are said to be {\bf \em coincident} 
if there exists a vertex $v$ such that either $e_1=(v,v_1), e_2=(v,v_2)$ or $e_1=(v_1,v), e_2=(v_2,v)$. 
We say that two edges $e_1, e_2$ are {\bf \em diverse} if $col(e_1) \cap col(e_2)=\emptyset$. Let us note that  coincident edges must be diverse.
By $mult(e)$ we denote the number of copies of the edge $e$ of $G$ in the multigraph $G_1$.
For a subset $E'$ of edges of $E$ by $mult(E')$ we denote $\sum_{e \in E'} mult(e)$.  By $\lambda(c)$ we denote the length of a cycle $c$.

\subsection{Preprocessing via alternating cycles} \label{preprocess}

Before building the multigraph $G_1$ we modify $C_1$ so that it differs from $C_{max}$ in a minimal way. An {\bf \em alternating cycle} in $C_{max} \oplus C_1$ is a set of edges that can be put in a sequence 
\newline $(v_1, v_2), (v_3, v_2), (v_3, v_4), (v_5, v_4), \ldots, (v_{k-1}, v_k), (v_1,v_k)$, 
in which edges belong alternately to $C_{max}\setminus C_1$ and $C_1 \setminus C_{max}$. By applying an alternating cycle $C_{alt}$ to $C_1$
we mean the operation, whose result is $C_1 \oplus C_{alt}$.  An alternating cycle $C_{alt}$ is {\bf \em good} if $C'_1=C_1 \oplus C_{alt}$ does not contain any  tricky $2$-cycle, tricky $3$-triangle or a tricky $2$-triangle of $R$, i.e., $C'_1$ is a relaxed cycle cover improving $C_{max}$.

\begin{fact} \label{altfact}
Let $C_{alt}$ be a good alternating cycle and $C'_1=C_1 \oplus C_{alt}$. Then $w(C'_1)\geq w(C_1)$.
\end{fact}
\dowod
Since $C_{max}$ is a maximum weight cycle cover of $G$, $w(C_{max} \oplus C_{alt}) \leq w(C_{max})$. Therefore, $w(C'_1) \geq w(C_1)$.
\koniec

We apply good alternating cycles to $C_1$ until it is no longer possible. We still call a new relaxed cycle cover $C_1$.
\begin{lemma} \label{alt}
After preprocessing, it holds that no alternating cycle is good.
\end{lemma}

\subsection{Path-coloring}

 Let $\Kd$ denote $\{i \in N: 1 \leq i \leq 20\}$, used for coloring $G^{20}_1$ and $\Kr =\{1,2,3,4\}$, used for coloring $G^4_1$. In this section $G_1$ denotes  $G^4_1$.
To {\bf \em path-color}  a multigraph $G_1$, or to {\bf \em path-color} it, means to  assign a color of $\Kr$ to each edge of  $G_1$ in such a way  that each color class consists of vertex-disjoint paths. Equivalently, we will be interested in path-coloring the underlying simple  graph $G_1$ so that each edge $e$ of $G_1$ is  assigned a subset $col(e)$ of $mult(e)$ colors of  $\Kr$ (and  each color class consists of vertex-disjoint paths).

 A path-coloring of $G_1$  will be carried out gradually. In the process each edge $e$  of $G_1$ can be either {\bf \em colored} - when it has $mult(e)$  colors assigned to it, or {\bf \em uncolored} - when it is   assigned no color.
A cycle $c$  is called {\bf \em monochromatic} if there exists a color $i$ of $\Kr$ such that each edge of $c$ is colored with  $i$ - $c$ is then a monochromatic cycle of color $i$.  Of course, a (partially) path-colored $G_1$  cannot contain any monochromatic cycle.
We will say that an edge $e$ is {\bf \em safe} if  no matter how we color the so far uncolored edges, it is guaranteed not to belong to any monochromatic cycle. For example suppose that $u$ has three incident edges in $G_1$ - $e_1=(u,v), e_2=(z,u), e_3=(z',u)$ such that $mult(e_1)=mult(e_2)=1, \ mult(e_3)=2$ and $col(e_1)=\{1\}$, $col(e_2)=\{2\}$ and $col(e_3)=\{3,4\}$. Also, $u$ has no other incoming edge in $G_1$. Then, clearly, $e_1$ is safe. By saying that an edge $e$ is {\bf \em $\bm{k}$-safe}  we will mean that $e$ is guaranteed not to belong to a monochromatic cycle of color $k$.
If $S$ denotes any subset of vertices of $G_1$, then $S^+$ denotes a set of edges $\{(u,v) \in G_1: u \in S, v \notin S\}$ and analogously,
$S^-=\{(u,v) \in G_1: u \notin S, v \in S\}$.
In path-coloring $G_1$ we are going to heavily use the following very helpful observation:

\begin{observation} \label{obs}
Suppose that an edge $e\in S^-$ is colored with  $k$ and no edge of $S^+$ is uncolored or colored with $k$. Then $e$ is $k$-safe.
Analogously, if an edge $e\in S^+$ is colored with  $k$ and no edge of $S^-$ is uncolored or colored with $k$, then $e$ is $k$-safe.
\end{observation}

  Suppose that an edge $e=(u,v) \in C_{max}$ is integral in $C_1$. Then it is called a {\bf \em  ray} if $u$ and $v$ belong to two different cycles of $C_1$, two different paths or a path and a cycle.  Otherwise, it is  called a {\bf \em  chord}. Note that a chord $e$ may also belong to $C_1$. A ray $r=(u,v)$  incident to a vertex on a cycle  or path $d$ of $C_1$  is said to be   a ray of $d$. If vertex $v$ belongs to $d$, then  $r$ is said to be an {\bf \em inray} of $d$. Otherwise, it is called its {\bf \em outray}.  An edge $e=(u,v) \in C_{max}$, which is  non-integral in $C_1$, is called a {\bf \em  border} (of a path of $C_1$).
	
Using Observation \ref{obs} we can apply the following simple method of coloring rays  of $C_1$.

\begin{lemma} \label{basekol}
Let $c$ be a cycle of $C_1$ such that each of its incident rays is  uncolored or safe. Then we are able to color all uncolored rays of  $c$ in such a way that each one of them is safe.
\end{lemma}
\dowod It is easy to guarantee that each newly colored ray is safe - it suffices if we  color inrays and outrays of $c$ with disjoint sets of colors, i.e., we partition $\Kr$ into $Z^-(c)$ and $Z^+(c)$ and each uncolored inray of $c$ is colored with a color of $Z^-(c)$ and each uncolored outray of $c$  with a color of $Z^+(c)$.
Then by Observation \ref{obs} and the fact that each previously colored ray is already safe, each ray of $c$ is safe. \koniec

Coloring rays so that they are safe does not mean, however, that there  always exists a possibility of coloring the remaining edges of $G_1$
so that we do not create a monochromatic cycle. Let us consider a few examples. 
If $c=(p,q,r,s)$ is a $4$-cycle of $C_1$ with $4$ inrays, each colored with $1$ and $4$ outrays, each colored with $2$, then
the only colors we can use on any edge of $c$ are $3$ and $4$.  Suppose now that we have a $4$-cycle $c=(p,q,r,s)$  
the same as above except for the fact that two of its outrays are also colored with $1$. 
It turns out that in this case we can path-color $c$, because two of its edges will be colored with $3$ and $4$ but one edge of $c$ can be colored with colors of $2$ and $3$ and one with $2$ and $4$.
 
We notice that a cycle $c$ of $C_1$ can be path-colored if it is possible for each color $k \in \Kr$ {\em not to be assigned} to at least one edge of $c$.
We define the flexibility of $e=(u,v) \in C_1$, denoted $flex(e)$, as $|col(e_1) \cap col(e_2)|$. (More generally, $flex(e)$ is defined as $b_{\alpha}-mult(e)-|col(e_1) \cup col(e_2)$, where $b_{\alpha}$ is the number of colors with which we are path-coloring a given multigraph.)
 For each cycle $c$ of $C_1$ we define its flexibility $flex(c)$ and  colorfulness $kol(c)$. The flexibility of $c$  is defined as $flex(c)=\sum_{e \in c} flex(e)$. Colorfulness $kol(c)$ denotes the number of colors of $\Kr$ used so far for coloring the edges of $G_1$ incident to $c$.  By $blanknk(c)$ we denote the number of uncolored edges of $C_{max}$ incident to $c$.
Using the above notions we define  the {\bf \em characteristic} $\chi(c)$ of a cycle $c$ of $C_1$ as follows: $\chi(c)=flex(c)+kol(c)+blank(c)$.
A cycle $c$ of $C_1$ is said to be {\bf \em blocked} if $\chi(c)<4$ and {\bf \em unblocked} otherwise.

\begin{lemma} \label{uzup}
Let $c$ be a cycle of $C_1$ that is not blocked and such that each of its incident rays is  colored and safe. Then we are able to color all edges and chords of  $c$ in such a way that each one of them is safe. 
\end{lemma}
\dowod

Let $e=(u,v)$ be any edge of $c$ and $e_1=(u,v'), e_2=(u',v)$ edges of $C'_{max}$ coincident with it. Then $e$ has to be colored with colors
of $\Kr \setminus (col(e_1) \cup col(e_2))$, i.e., it cannot be colored with any color assigned to the edge coincident with it.

Suppose first that $c$ does not contain any chords. If each color $k \in \Kr$ is assigned to some ray  of $c$, then we are already done. By coloring each edge $e$ of $c$ with any $mult(e)$ colors of $\Kr$ that are not assigned to any edges of $C_{max}$ coincident with $e$, we achieve that each color $k \in \Kr$ is {\em not} assigned to some edge  of $c$, thus $c$ is not monochromatic with respect to any color of $\Kr$. Otherwise, if $kol(c) < 4$ and thus not every color $k \in \Kr$ is assigned to some ray  of $c$, we still have that $\chi(c) = kol(c)+flex(c) \geq 4$. Therefore, $flex(c)=\sum_{e \in c} flex(e) \geq 4-kol(c)$. This means that for each edge $e$ of $c$ with $flex(e)>0$ we can choose $1$ color of $\Kr$,  which will {\em not} appear on $e$.  Recall that $e$ is colored with $mult(e)=2$ colors and $flex(e)=|col(e_1) \cap col(e_2)|$, where $e_1, e_2$ are edges coincident with $e$. This way we can distribute all colors of $\Kr$ not assigned to any rays of $c$ among edges of $c$ and ensure that each color of $\Kr$,  which is not assigned to any ray of $c$, does not appear on some edge of $c$. Hence, $c$ will again not be monochromatic w.r.t. any color of $\Kr$. The remaining part of the proof, when $c$ may contain chords, is in Section \ref{miss}. \koniec

Similarly, as cycles of $C_1$ may be blocked, paths of $C_1$ can become non-path-$20$-colorable too. Or, more precisely,  halfy $2$-cycles 
of $C_1$ can become blocked.  Let $p_1=(u_1, \ldots, u_k)$ denote a path of $C_1$. Then both $u_1$ and $u_k$ belong to two different halfy $2$-cycles $c_1=(u_1,v_1)$ and $c_2=(u_k, w_l)$ of $C_1$. Thus $C_1$ contains also paths $p_2=(v_1, \ldots, v_{k'})$ and $p_3=(w_1, \ldots, w_l)$, though it may happen that $p_2=p_3$. If $p_1$ consists of more than one edge, then $C_{max}$
contains edges $a_1=(u'_2,u_2), a'=(u_{k-1},u'')$, none of which is a border. Each of these edges is called an {\bf \em antenna} (of  $p_1$).  
$a_1$ is also said to be  an antenna of $c_1$ and $a'$ of $c_2$.

\begin{figure}[h]
\centering{\includegraphics[scale=0.8]{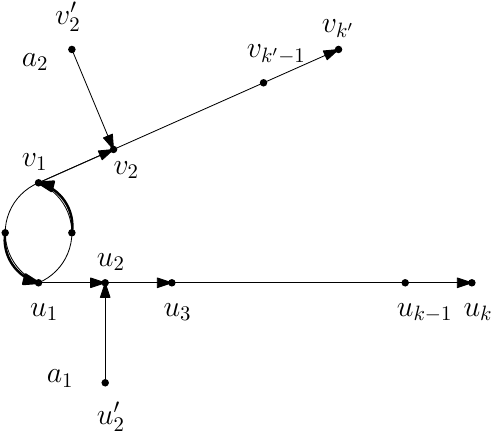}}
\caption{{\scriptsize  Antennas $a_1,a_2$ of a halfy $2$-cycle $(u_1, v_1)$.}
} \label{antennas}
\end{figure}

\begin{fact} \label{anteny}
Let $c=(u_1,v_1)$ be a halfy $2$-cycle of $C_1$ with two antennas $a_1, a_2$. Then, in any path-coloring of $G_1$ the antennas $a_1$ and $a_2$ have to be diverse.
\end{fact}  
\dowod Suppose that $C_1$ contains paths $p_1=(u_1,u_2, \ldots, u_k)$ and $p_2=(v_1,v_2, \ldots, v_l)$. Then the antennas $a_1$ and $a_2$
have the form $a_1=(u'_2,u_2)$ and $a_2=(v'_2,v_2)$, see Figure \ref{antennas}.
Since $c$ is a $2$-cycle, its edges have to be diverse. Also, we may notice that $(u_1,u_2)$ has to be colored in the same way as $(v_1,u_1)$
and $(v_1,v_2)$ in the same way as $(u_1,v_1)$. Therefore $(u_1,u_2)$ and $(v_1,v_2)$ have to be diverse. Also $col(u_1,u_2) \cup col(v_1,v_2)=\Kr$.
Since $a_1$ and $(u_1,u_2)$ have to be diverse and so do $a_2$ and $(v_1,v_2)$, $a_1$ and $a_2$ have to be diverse as well.\koniec

If $a_1, a_2$ are two antennas of a $2$-cycle $c$ and $a_1$ is already colored but $a_2$ not, then we say that color $col(a_1)$
is {\bf \em forbidden} on $a_2$. A halfy $2$-cycle $c$ of $C_1$ is said to be {\bf \em blocked}, if it has two antennas and they are not diverse. 

Multigraph $G_1$ is {\bf \em blocked} if at least one cycle or halfy $2$-cycle of $C_1$ is blocked and {\bf \em unblocked} otherwise. $G_1$ is {\bf \em safe} if each of its  colored edges is safe.
We say that a cycle or path of $C_1$ is {\bf \em unprocessed} if at least one of its rays is uncolored.
To {\bf \em process} a cycle/path of $C_1$ means to color its rays  so that $G_1$ is safe and unblocked, assuming that it was safe and unblocked before.

We are now ready to state the algorithm for path-$4$-coloring $G_1$.

\begin{algorithm}[H]
  \begin{algorithmic}[1]
    \While {there there exists an unprocessed cycle of $C_1$} 
		\State $c \leftarrow$  an unprocessed cycle of $C_1$ with a minimal  number of  uncolored rays
		\State process $c$
		\EndWhile
		\While {there there exists an unprocessed path of $C_1$} 
		\State $p \leftarrow$  any unprocessed path of $C_1$ 
		\State process $p$
		\EndWhile
		
		\State color the remaining uncolored edges  in such a way that each one of them is safe
    
  \end{algorithmic}
	\caption{Path-coloring of $G^{4}_1$ without halfy triangles}
\label{alg:path4color}
\end{algorithm}


In what follows we prove the correctness of Algorithm \ref{alg:path4color}
We say that two rays $r_1 =(u, u'), r_2=(v',v)$ of a $2$-cycle $c$ are {\bf \em complementary (on $c$)}  if one of the edges  $(u,v),(v',u')$ belongs to $c$.

\begin{fact}
 A $2$-cycle is blocked only if some two of its non-complementary rays $r_1, r_2$ are not diverse.
\end{fact}


\begin{lemma} \label{ost}
Let $c$ be any cycle of $C_1$ with an uncolored ray $r$ coincident with an edge $e$ of $c$. Let $r'$ denote another edge of $C_{max}$ coincident with $e$.
Then by coloring $r$ with any color $k \notin  col(c)$ we increase $kol(c)$ by $1$.
If $r'$ is already colored with $k'$, then by coloring $r$ with $k'$, we increase $flex(c)$ by $1$.
Thus the set of colors $Z(r)$ each of which can be applied on $r$ to increase $kol(c)+flex(c)$ has size $4-kol(c)+|col(r')|$.

\end{lemma}

\begin{lemma} \label{min}
Let $c$  be an unprocessed cycle of $C_1$ that at some step of Algorithm \ref{alg:path4color} has a minimal number of  uncolored rays. It is always possible to process $c$.
\end{lemma}

\dowod We divide the set of colors $\Kr$ into two sets $Z^+(c)$ and $Z^-(c)$. Next, we color each uncolored inray of $c$ with one of the colors of $Z^-(c)$ and each uncolored outray of $c$ with one of the colors of $Z^+(c)$. This way each newly colored ray is safe - by Observation \ref{obs} and the assumption that all previously colored rays are safe. Now, we  prove that we can carry out the above in such a way that no cycle or halfy $2$-cycle of $C_1$ becomes blocked. Let $O_2$ denote a set of cycles of $C_1$ such that each cycle $c$ of $O_2$ has no chords and has exactly two uncolored rays coincident with the same edge $e$ of $c$.
\begin{claim} 
It never happens that each of the uncolored rays of a cycle of $O_2$ is incident to two cycles of $O_2$.
\end{claim}
\dowod
Suppose to the contrary that each of the uncolored rays of $O_2$ is incident to two cycles of $O_2$. Then the rays of these cycles together
with edges of these cycles, with which the uncolored rays are coincident form a good alternating cycle contradicting Lemma \ref{alt}. \koniec

By this claim, we can always choose for processing a cycle that  does not belong to $O_2$.
Let $Z(c,e)$ denote the set of colors that can be used for coloring ray $e$ of $c$ so that $c$ remains unblocked. For simplicity assume also that $c$ has no chord.  Chords make path-coloring easier. By Lemma \ref{ost}, we have:

\begin{itemize}
\item  If $blank(c)=1$, then $|Z(c,e)| \geq 2$ for the uncolored ray $e$ of $c$.
\item If $blank(c)=2$ and $c \notin O_2$, then $|Z(c,e)| \geq 3$ for at least one uncolored ray $e$ of $c$.
\item If $blank(c)=2$ and $c \in O_2$, then $|Z(c,e)| \geq 2$ for each uncolored  ray $e$ of $c$.
\item If $blank(c)=3$, then $|Z(c,e)| \geq 4$ for at least one uncolored ray $e$ of $c$.
\end{itemize}

Let $Z(c)=\bigcup_{e} Z(c,e)$. Suppose first that $c$ has exactly one uncolored ray $r$.   If $r$ is incident to another cycle $c'$ of $C_1$  then 
we also have to guarantee that $c'$ does not become blocked.  We need to concern ourselves with blocking $c'$ only if $blank(c') \leq 2$. If $Z(c) \cap Z(c') \neq \emptyset$, then we color $r$ with a color $k \in Z(c) \cap Z(c')$.
Otherwise, we color $r$ with a color $5$. We may be forced to use a color $5$ only if (i) $blank(c') = 2$ and $c' \in O_2$ or (ii) $blank(c') = 1$. 
If we do have to use color $5$, then we can still easily obtain a path-$20$-coloring of $G^{20}_1$ because color $5$ will be changed into the set $\{5,10,15,20\}$ in $G^{20}_1$. Generally, the transfer from $G^{4}_1$ to $G^{20}_1$ consists in the following.
If an edge $e \in C_1$  is colored with  $i$ of $\Kr$ in $G^4_{1}$, them we assign the subset $\{5i, 5i-1, 5i-2, 5i-3, 5i-4\}$ of $\Kd$ to $e$ in $G^{20}_1$.
If some edge $e \in C_{max}$ in $G^4_{1}$  is colored with $i$, then we assign it the set $\{5i-1, 5i-2, 5i-3, 5i-4\}$ in $G^{20}_1$.
Additionally, if some edge $e \in C_{max}$ in $G^4_{1}$ is colored with $5$, then in the coloring of the edges of $C_1$ that belong to cycles of $C_1$ incident to $e$ we replace  colors in $\{5,10,15,20\}$  with the other colors of the same set, for example $5$ with $15$ and $10$ with $20$.

If $c$ has exactly two uncolored rays $r_1, r_2$, then we may assume that $c \notin O_2$. 
Suppose first that $r_1, r_2$ are also rays of another cycle $c'$ of $C_{max}$. Since $|Z(c)| \geq 3$ and $|Z(c')| \geq 2$, at least one color  belongs to $Z(c) \cap Z(c')$. We apply $\min\{2, |Z(c) \cap Z(c')|\}$ colors on $r_1, r_2$ and in the case when $Z(c) \cap Z(c')$ contains only one color, we also use color $5$.
If $r_1, r_2$ are rays of two different cycles $c',c''$, then the same argument goes through, because it may happen that $Z(c')=Z(c'')$.

If $c$ has  $3$ or more uncolored rays, then the proof is similar but in such cases we never have to apply color $5$.
Let us also remark that a ray $e$ of $c$ may be also an antenna of some halfy $2$-cycle $c'$ and hence one color may be forbidden on it (because it needs to be diverse with the second antenna of $c'$) but such cases can be treated in the same way as though $e$ was a ray of a $2$-cycle $c'$ with exactly one colored ray - then also one color is forbidden on $e$. Also, we can avoid cases when each ray of a $2$-cycle has exactly the same color forbidden, for example by forbidding that color on that antenna which is colored last.\koniec

If $a$ is an antenna of two different halfy $2$-cycles, then it is said to be a {\bf \em bilateral antenna}.

\begin{lemma} \label{path}
Let $p$ be an unprocessed path  of  $C_1$. Then it is possible to process it.
\end{lemma}
\dowod The proof is similar to the one above. Since paths are processed after cycles of $C_1$, the only thing we have to take care of
is that antennas of the same halfy $2$-cycle are diverse. The path $p$ has at most two incident bilateral antennas $a_1, a_2$ and if it does, then at most one of them is an inray and at most one an outray of $p$. Assume that $a_1$ is an inray and $a_2$ an outray. (They may also be chords.) Each $a_i$ may have to be diverse with two different edges. Thus for each $a_i$ it may happen that up to $2$ colors are forbidden on it.  Let $Z_i$ denote the set of colors forbidden on $a_i$.  We  partition $\Kr$  into
two $2$-element sets $Z^-(p)$ and $Z^+(p)$ so that $|Z^-(p) \setminus Z_1| \geq 1$ and $|Z^+(p) \setminus Z_2| \geq 1$. 
To achieve this we  divide each of $Z_1 \cap Z_2$ and $\Kr \setminus (Z_1 \cup Z_2)$ (almost) equally between $Z^-(p)$ and $Z^+(p)$.  Also, $Z_1\setminus Z_2$ goes to $Z^-(p)$ and $Z_2 \setminus Z_1$ to $Z^+(p)$.    Since  $|Z_1 \cap Z_2| \leq 2$, it means that  $Z^-(p)$ (resp. $Z^+(p)$) contains at most $1$ color of $Z_1$ (corr $Z_2$) as required.
Then we are able to color each ray of $p$ so as to ensure that each antenna is diverse with the required antennas.
\koniec

\section{Path-coloring in the presence of  halfy triangles} \label{pathcolmul}

Recall that a tricky triangle $t=(p,q,r)$ is {\bf \em halfy} if $C_1$ contains exactly one half-edge of some edge of $t$ or of $opp(t)$.
Each halfy triangle has either (i) three incoming paths of $C_1$ and one outgoing or symmetrically, three outgoing paths of $C_1$ and one incoming or
(ii) two incoming paths of $C_1$  or symmetrically, two outgoing paths of $C_1$.
We now describe the number of copies in $G^4_1$ of each edge of $t \cup opp(t)$ for a halfy triangle $t$. An edge $e \notin C_1$, such that $G^4_1$ contains exactly $2$ copies of $e$ is called a {\bf \em b-edge}.

\begin{figure}[h]
\centering{\includegraphics[scale=0.75]{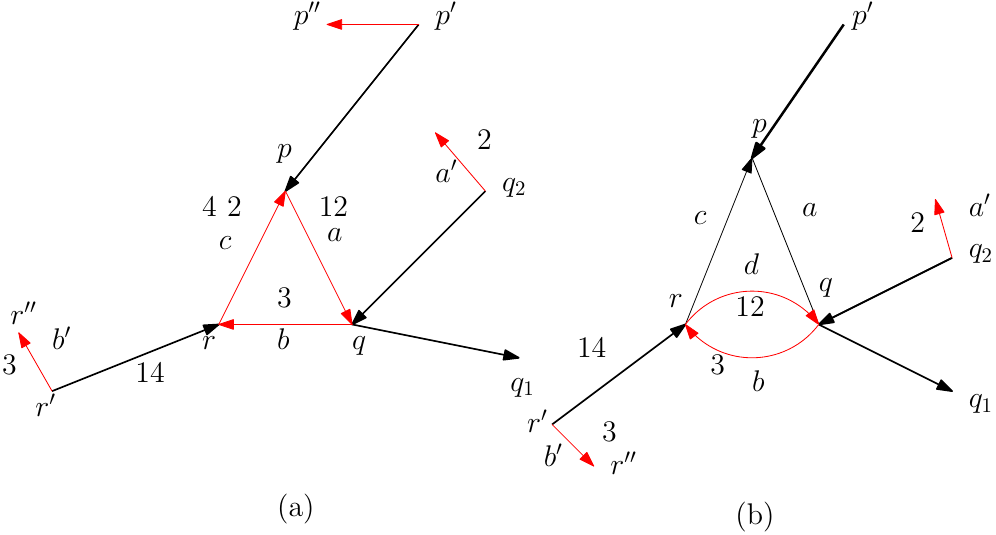}}
\caption{  Halfy triangles with $3$ incoming and $1$ outgoing edges of $C_1$: a tricky $3$- and $2$-triangle $t=(p; q,r)$. In both cases the inherited color on the main b-edge is $2$.}
 \label{htrgscolor}
\end{figure}

Let $t=(p,q,r)$ be a halfy triangle consisting of edges $a=(p,q), b=(q,r), c=(r,p)$. If $t$ is a $2$-triangle, then $(r,q)$ is a $2$-cycle of $C_{max}$ and we denote $d=(r,q)$.
Suppose first that $C_1$ contains edges $(p', p), (q_2,q), (q,q_1), (r', r), (p', p'')$, see Figure \ref{htrgscolor}.  
If $t$ is a $3$-triangle, $G^4_1$ contains $2$ copies of each of $a,c$ and $1$ copy of $b$. We call $b$ an {\bf \em s-edge} of $t$, $a$ the {\bf \em main b-edge} of $t$ and $c$ the {\bf \em secondary b-edge} of $t$. If $t$ is a $2$-triangle, then $G^4_1$ contains $2$ copies of $d=(r,q)$ and $1$ copy of $b$ and $0$ copies of $a$ and $c$.  We call $d$ the {\bf \em main b-edge} of $t$ and $b$ an {\bf \em s-edge} of $t$.  
In each of these two cases the main b-edge of $t$ is called an {\bf \em inner antenna} of $t$.  An edge $(r', r'')\in C_{max}$ is an {\bf \em outer antenna} of $t$ if $t$ is a $2$-triangle or $t$ is incident to a cycle of $C_1$. In the other case - when $t$ is a $3$-triangle not incident to a cycle of $C_1$, the edge $(p', p'')$ is its {\bf \em outer antenna}.
Suppose next that $C_1$ contains edges $(p',p), (q',q)$ and $t$ has no outgoing edge in $C_1$, see Figure \ref{htrgs2}. If $t$ is a $3$-triangle, $G^4_1$ contains $2$ copies of each of $a,c$ and $3$ copies of $b$. If $t$ is a $2$-triangle, $G^4_1$ contains $1$ copy of each of $a,c,d$ and $3$ copies of $b$. Each of the edges $(p', p''), (q', q'')$ is called 
an {\bf \em antenna} of $t$. 
For each halfy triangle $t$, we require that both antennas of $t$ are diverse - otherwise it is {\bf \em blocked}.

If the cycle cover $C_1$ contains halfy triangles, then in coloring $G_1$, we would still like to take advantage of Observation \ref{obs}  and color all rays of a given cycle or path in one step. Here, however, this task is  more complicated because of b-edges. Consider an edge $e=(u,v)$ belonging to some path or cycle  of $C_1$. Suppose that it is coincident with two rays
$r=(u,u'), r'=(v',v)$ of this path or cycle. If $r$ is a b-edge, meaning that $mult(r)=2$,  then it is impossible to color $r$ and $r'$ with disjoint sets of colors of $\Kr$. This is because, both $r$ and $r'$ have to be diverse with $e$ and hence $col(r) \cup col(e)=\Kr$, which means that $col(r') \subseteq col(r)$. To deal efficiently with the coloring of b-edges, we divide the two colors of $col(r)$ assigned to any b-edge
$r$ into $r$'s {\bf \em own color}, denoted $col'(r)$ and a {\bf \em color inherited} by $r$, denoted $col''(r)$. If in the above example,
$r'$ is not a b-edge, then a color inherited by $r$ satisfies $col''(r)=col(r')$. On the other hand, if $r'$ is a b-edge, then $col''(r)=col'(r')$.
We call $r'$ an {\bf \em ally} of $r$ and denote as $al(r)=r'$. Analogously, if $r'$ is a b-edge, then $r$ is called an ally of $r'$.
 For every b-edge, its inherited color comes from its {\bf \em ally}.
If both $r$ and $r'$ are b-edges, then it holds that
$col(r)=col(r')$ and $col'(r)=col''(r'), \ col'(r')=col''(r)$. If $r$ is not a b-edge, then $col'(r)$ denotes simply $col(r)$.
We say that two antennas $a_1, a_2$ of a halfy cycle $c$ of $C_1$ are {\bf \em diverse} if the sets of its own colors are disjoint, i.e., if $col'(a_1) \cap col'(a_2)=\emptyset$. 




Let $t=(p,q,r)$ be a halfy $3$-triangle from above with three incoming paths, see Figure \ref{htrgscolor} (a), and let $a'$ denote the ally of $a$ and $b'$ the edge $(r', r'')$.
In order  to guarantee that $a$ is safe with respect to its inherited color  $ col''(a)$, we proceed as follows. Let $k=col''(a)$. If $k$ is not already assigned to $c$,
we can assign it to $b$ and be done - we can notice that if we do not assign $k$ to $col'(c)$, the edge $a$ is safe w.r.t. $k$, because the only cycle the path consisting of $a$ and $b$ is contained in is $t$.  If $k\in col(b')$, then we notice that the edge $(r',r)$ will not be colored with $k$, therefore $a$ and $c$ are safe w.r.t $k$. Otherwise, i.e, if $k \in col''(c)$ and $k \notin col(b')$, we {\bf \em shadow} $b$, i.e., we assign $col'(b')$ to $b$ and forbid the color $k$ on the edge $(r', r)$ - we are able to do this because $flex(r', r)=1$. Also, since we require that $col'(b') \neq col'(a)$,  $t$ will not be monochromatic w.r.t. $col'(b')$. In Figure \ref{htrgscolor} the inherited color $k=2$. 

Suppose next that $t=(p;q,r)$ is a halfy $2$-triangle from above, see Figure \ref{htrgscolor} (b).  
We assign $k=col'(a')$ to $d$. If $k \in col'(b')$, the edge $(r',r)$ will not be colored with $k$, therefore $d$ is safe w.r.t. $k$. Otherwise, we {\bf \em shadow} $b$ - we assign $col'(b')$ to $b$ and forbid $k$ on $(r',r)$. 

\subsection{Algorithm}

When coloring rays of a cycle or path $s$ of $C_1$, we may not be able to color b-edges and s-edges incident to  $s$ fully, because their allies have not been colored yet.   For this reason, we introduce the notion of precoloring. To {\bf \em precolor} an edge $r$  means to assign it its own color $col'(r)$ if $r$ is not a secondary b-edge or an s-edge and leave uncolored otherwise. 

Below we show that we can guarantee that each ray of a given cycle $c$ is safe by using a similar approach as previously, where we colored 
inrays and outrays with disjoint sets of colors. The modification consists in the fact that for b-edges, we only require that colors owned by them, i.e., sets $col'(r)$ obey this partition.

We say that an edge is {\bf \em conditionally safe} if it is guaranteed not to belong to a monochromatic cycle under the condition
that all halfy triangles are colored in the manner described above.
We say that $G_1$ is blocked if there exists a cycle or a halfy cycle  of $C_1$ that is blocked. Otherwise, $G_1$ is unblocked.
To {\em \bf process} a cycle or path $s$ of $C_1$ means to precolor all its rays in such a way that all of them are conditionally safe
and $G_1$ is unblocked. 

\begin{algorithm}[H]
  \begin{algorithmic}[1]
    \While {there there exists an unprocessed cycle of $C_1$ without any incident b-edge} 
		\State $c \leftarrow$  an unprocessed cycle of $C_1$ with a minimal  number of  uncolored rays
		\State process $c$
		\EndWhile
		\State process all cycles of $C_1$ with an incident b-edge
		\While {there there exists an unprocessed path of $C_1$} 
		\State $p \leftarrow$  any unprocessed path of $C_1$ 
		\State process $p$
		\EndWhile
		\State color fully all b-edges and s-edges
		
		\State color the remaining uncolored edges  in such a way that each one of them is safe
    
  \end{algorithmic}
	\caption{Path-coloring of $G^{4}_1$ with halfy triangles}
\label{alg:path4colorhalfy}
\end{algorithm}


\begin{lemma} \label{bcycle}
Let $c$ be an unprocessed cycle of $C_1$ that has an incident b-ray.  Then it is always possible to process $c$.
\end{lemma}

\dowod  We show that we can always ensure that $c$ is not blocked by  coloring the rays of $c$ in such a way that $\chi(c) \geq 4$. 
To process $c$, we need to color all rays of $c$, which are not s-edges. We partition $\Kr$ into two disjoint sets $Z^+(c)$ and $Z^-(c)$. If we  precolor each uncolored outray with colors of $Z^+(c)$ and each uncolored inray with colors of $Z^-(c)$, then each such newly colored ray becomes safe. As for s-edges, we have more freedom in coloring them and do not have to observe this partition.  Recall that each s-edge $e$ is guaranteed to be safe, of course as long as it is not assigned the same  color $k$ as the other b-edge(s) of the same halfy triangle.

While coloring the rays of $c$, we also have to ensure that no other cycle or halfy cycle of $C_1$ becomes blocked.  We do not need to concern ourselves with blocking cycles of $C_1$ different from $c$, because cycles of $C_1$ with no incident b-rays are already processed and by the current lemma we are always capable of processing a cycle of $C_1$ with an incident b-ray. Thus  we only have to take care of halfy cycles.
Since each ray of $c$ is an antenna of at most one halfy cycle,  every ray of $c$ has to be diverse with at most one edge.

First, we notice that if the number of uncolored rays of $c$, which are not s-edges is at least $4$, then we can easily
guarantee that $kol(c) \geq 4$ by assigning different colors to $col'(r)$ of each uncolored ray which is not an s-edge. 
More generally, we can easily guarantee that $kol(c)+flex^+(c) \geq 4$ whenever $kol(c)+flex^+(c)+ blank(c)- |\{ $\ s-edges incident to \ $ c \}|  \geq 4$. We can do that by using  $4-kol(c)-flex^+(c)$ new colors (not already included in $col(c)$) on the uncolored rays, which are not s-edges.

Let us next observe that in the cases  when all uncolored rays of $c$, which are not s-edges, are either all outrays or all
inrays, all rays of $c$ are  safe under the obvious condition that each b-ray $r$ of $c$ is diverse with some edge of the halfy cycle containing $r$.  Thus we do not have to shadow  s-edges incident to $c$ and can use new colors on them. This means that in such cases we are always able to use
$blank(c) \geq 4-kol(c)-flex(c)$ new colors on uncolored rays of $c$.
Let us also notice that if $c$ has an edge $e$ with two rays coincident with it, none of which is a b-ray or an s-edge, then $kol(c)+flex(c)\geq 2$ and we can use two new colors on the remaining b-rays, s-edges and rays and be done. (If $c$ has at least one incident b-edge, then at the moment of processing it, $blank(c) \geq 2$,
because apart from this b-edge, an s-edge is also incident to $c$.)

Also, as long as $c$ has at least two previously colored rays, we can easily ensure that $kol(c)+flex(c) \geq 2$.

We are thus left with the following two cases: (i) $c$ is a $2$-cycle with one b-inray and one b-outray  
or (ii) $c$ is a triangle with one b-inray and two outrays or vice versa with one b-outray and two b-inrays.

\begin{figure}[h]
\centering{\includegraphics[scale=0.9]{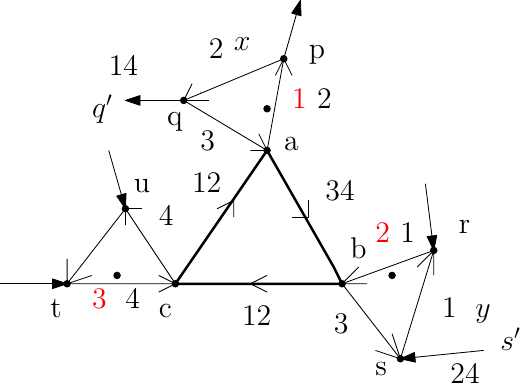}}
\caption{ A triangle $(a,b,c)$ of $C_1$ has three incident b-rays: $(a,p), (r,b), (t,c)$
(denoted by a dot) contained in three halfy $3$-triangles.
The own colors of $(a,p), (r,b), (t,c)$  are $1,2$ and $3$ correspondingly, i.e, $col'(a,p)=1, col'(r,b)=2, col'(t,c)=3$.  Thus $col''(a,p)=2, col''(r,b)=1$. To make the color $2$ safe on $(a,p)$, we assign it also to $(p,q)$ and shadow, which means that $2$ will not appear on $(q,q')$. Similarly we shadow $1$ on the triangle $(r,b,s)$. Since neither $1$ nor $2$ will appear on $(b,s)$, in the case when $(b,s)$ is colored $3$, the edge $(t,c)$ is free to choose any inherited color it likes (we should of course choose a color different from $3$, because $3$ is already assigned to it and choosing $1$ or $2$ would be a bad idea because then $\chi(a,b,c)=3$ and $(a,b,c)$ would be blocked.) We do not have to shadow $4$, we may simply assign it to $(c,u)$ thus making it safe.}
 \label{Btriangle}
\end{figure}

In case (ii) we notice that one edge $e$ of $c$ is coincident with a b-inray of $c_1$ and a b-outray of $c_2$. Suppose that the third b-ray of $c$ belongs to a halfy cycle $c_3$. We assign three different colors to the b-rays, $k_i$ to $c_i$ for $i=1,2,3$.  We notice that we need to do the shadowing only for $k_1$ and $k_2$. Since the ally of the main $b$-edge of $c_3$ is the s-edge of either $c_1$ or $c_2$, this $s$-edge will be not be assigned $k_1$ or $k_2$. This means that it will be assigned either the fourth possible color (and then $kol(c)=4$) or it will be assigned $k_3$, and in this case the inherited color of the main b-edge of $c$ may be chosen arbitrarily and then we of course choose the fourth color again. An example is depicted in Figure \ref{Btriangle}.

\begin{fact}
\begin{enumerate}
\item Each edge of $C'_{max}$ is an antenna of at most two halfy cycles of $C_1$.
\item If an edge $e \in C'_{max}$ is an antenna of two halfy cycles, then it is not incident to a cycle of $C_1$.
\end{enumerate}
\end{fact}

\begin{lemma} \label{path1}
It is always possible to process a path of $C_1$.
\end{lemma}
\dowod
 Since paths are processed after cycles of $C_1$, only halfy cycles of $C_1$ can become blocked. To prevent this, we have to ensure  that antennas of the same halfy cycle are diverse or weakly diverse. The path $p$ has two outer  antennas $a_1, a_2$ and at most one of them is an inray and at most one the outray of $p$. (Each one of them may also be a chord of $p$.)  Assume that $a_1$ is an inray and $a_2$ the outray. Each $a_i$ may have to be diverse with two different antennas, i.e., may be bilateral.  Note that no other ray of $p$ is a bilateral antenna. The path $p$ may also have one or two weak antennas $a'_i$. An uncolored weak antenna $a'_i$ has to be diverse with $a_i$. It may also be an outer antenna of some other path $p'$.
Let $Z_i$ and $Z'_i$ denote the set of colors forbidden on $a_i$ and $a'_i$, respectively. Let us note that if $a_i$ is uncolored, then $|Z'_i| \leq 1$. If $a'_i$ is an uncolored inner antenna, then $|Z_i| \leq 1$. If $a'_i$ is a weak antenna (or if $p$ has no antenna $a'_i$) , then $|Z_i|$ may be equal to $2$.

For processing $p$ we always choose a path that has at most one uncolored weak antenna, suppose that it is $a'_1$. Since each halfy triangle has at most one weak antenna, the average number of weak antennas per path is at most one. Therefore, we can indeed always choose such a path.
If $a_1$ is colored, then the number of colors forbidden on $a'_1$ may be two.
Assume that none of the three antennas is already colored, as the other cases are contained in this one.
Since $a_1$ and $a'_1$  have to be diverse, we need two colors for coloring them. Therefore, $Z^-(c)$ needs to satisfy: $|Z^-(c)\setminus Z_1 \cup  Z^-(c)\setminus Z'_1| =|Z^-(c) \setminus (Z_1 \cap Z'_1) \geq 2$. The set $Z^+(c)$ needs to satisfy $|Z^+(c) \setminus Z_2| \geq 1$.
We can achieve this because the worst case can be described as $Z_1=Z_2=\{1,2\}$ and $Z'_1=\{1\}$. Then we set $Z^-(c)=\{2,3\}$ and $Z^+(c)=\{1,4\}$ and are done.
After computing such a partition, we are able to color each ray of $p$ so as to ensure that each antenna is  diverse with the required antennas. \koniec

\subsection{Full coloring of b-edges and s-edges}

During the processing of paths and cycles of $C_1$ we do not assign color $col'(e)$ for any edge $e$, which is either a secondary b-edge  or an s-edge. As a result some b-edges cannot be assigned their inherited color $col''(e)$. This happens for any b-edge, whose ally is an s-edge or a secondary b-edge.

To be able to complete the process of coloring all b-edges and s-edges, we introduce a directed graph $D=(V_D, E_D)$, which shows the dependencies between halfy triangles containing such edges. The vertex set of $D$ consists of all halfy triangles.  The edge set $E_D$ contains an edge $(t, t')$ iff the  ally of the main b-edge of $t'$
is either an s-edge of $t$ or a secondary b-edge of $t$. The direction of an edge $(t, t')$ reflects the fact that $t$ needs to be fully colored to be able to complete the coloring of $t'$. Note that each vertex of $V_D$ has at most one incoming edge and at most two outgoing edges. To {\bf \em D-process} a directed path or cycle $s$ of $D$ means to complete the coloring of each halfy triangle $t$ corresponding to any vertex on $s$ in such a way that $G_1$ remains unblocked.

Notice that any two cycles of $D$ are vertex-disjoint.

\begin{lemma}
It is possible to D-process each cycle of $D$.
\end{lemma}
\dowod
 The general  idea is that we can choose more or less an arbitrary triangle $t$ on $c$, color it fully and in this way trigger the coloring of the remaining triangles on $c$. We will show that this approach will not cause difficulties and that in each case we will be able to fully color triangles on $t$ in the ''safe manner'' and thus to D-process $c$.
The coloring of the main b-edge of each triangle $t$ on $c$ depends either on an s-edge or the secondary b-edge of the triangle $pred(t)$, which is its predecessor. 
Suppose that the order of triangles on $c$ is $(t_0, t_1, \ldots, t_k)$.

If there exists a triangle $t$ on $c$ such that the coloring of its successor $t'=succ(t)$ depends on the secondary b-edge $e_2(t)$ of $t$, then $col''(e_2(t) \neq col'(e_1(t))$ and we assign to $col'(e_2(t))$ any color different from these two. We can notice that when we color fully all main edges of the triangles on $c$ except for $t$, we will be able to complete the coloring of $t$ regardless of which color will be assigned to $col''(e_1(t))$. 

\begin{figure}[h]
\centering{\includegraphics[scale=0.75]{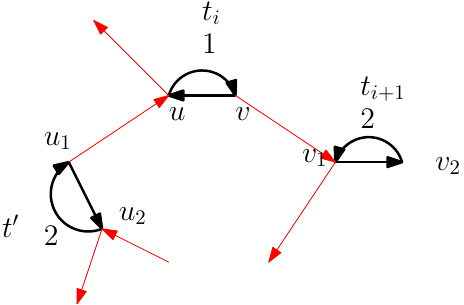}}
\caption{The own color $2$ of the main b-edge $(v_2, v_1)$ of a $2$-triangle $t_{i+1}$ is forbidden on the s-edge $(u,v)$ of a $2$-triangle $t_i$, because this is an antenna of $t'$.}
 \label{dependforbid}
\end{figure}

If there exists a triangle $t_i$ that corresponds to a halfy $2$-triangle (hence $t_i$ has only one b-edge) and such that the main color of $e_1(t_{i+1})$ is forbidden on the s-edge $s(t_i)$ of $t_i$, we start from its successor $t_{i+1}$ and assign it any color not forbidden on $s(t_i)$ and of course different from $col'(e_1(t_{i+1})$. The same color must then be assigned to $s(t_i)$. When we return to $t_i$ (after coloring all triangles on $c$ except for $t_i$), its b-edge  will be assigned some inherited color $k'$ but we do not have to shadow $k'$ - it will be made safe by the triangle $t'$ that forbids the color on the s-edge of $t_i$ - $col'(e_1(t'))$ is  the same as   $col'(e_1(t_{i+1}))$. The reason is the following. See Figure \ref{dependforbid}. Any cycle $c(t_i)$ in $G_1$ of length greater than $2$ that contains the main b-edge of $t_i - e_1(t_i)=(u,v)$, apart from a $2$-cycle consisting of the main b-edge and the s-edge of $t_i$, also has got to contain the main b-edge of $t'$. Thus, if $c(t_i)$ were to be monochromatic, it would have to have either the own color of $e_1(t')$ or its inherited color. However, for $t'$ we perform the shadowing to ensure that we do not create a monochromatic cycle of color $col''(e_1(t'))$. As for $col'(e_1(t'))$ we can notice that $c(t_i)$ also contains the edge $e'=(v,v_1)$ of $C_1$ connecting $t_i$ and $t_{i+1}$ and this edge is coincident with $e_1(t_{i+1})$ and since $col'(e_1(t'))=col'(e_1(t_{i+1})$, the color $col'(e_1(t'))$ cannot occur on $e'$.

If there exists a triangle $t$ on $c$ such that the coloring of its successor $t'$ depends on the s-edge $s(t)$ of $t$, we choose a color $k$ not occurring on any edge of $t$ and different from $col'(e_1(pred(t))$ and assign it both to $e_1(t)$ and $s(t)$. Further while coloring the main edges of other triangles on $c$ we use one of the two colors of $e(t)$ for the inherited color of any main b-edge of a triangle on $c$.
This way when we come to $t'=pred(t)$ its inherited color will be one of the two colors occurring on $e_1(t)$ - thus it will be already shadowed.


Otherwise all b-edges of all triangles on $c$ are colored with the same color, $k$, and the ally of each main b-edge is the secondary b-edge of its successor.
In this case we can deal with each triangle separately, by assigning the same color to $e_1(t)$ and $s(t)$.

 \koniec

\section{Tricky cycles}\label{trickysec}

Let $c$  be any of the following cycles:
\begin{enumerate}
\item a $2$-cycle that either belongs to $C_{max}$ or is a subcycle of a triangle of $C_{max}$.
\item a $3$-triangle.
\item a $2$-triangle. 
\end{enumerate}

The occurrence of any such cycle in $C_1$ makes $G^{20}_1$ non-path-$20$-colorable. However, it is sometimes possible to make a subgraph of $G^{20}_1$ induced on $c$ (and possibly its neighboring edges) path-colorable via local replacements. In cases where $c$ is a $2$-cycle of $C_{\max}$, a $3$-triangle, or a $2$-triangle, these replacements may involve only edges with both endpoints on $c$. If $c$ is a $2$-cycle that forms a subcycle of a triangle in $C_{\max}$, we may also use edges of $C_{\max}$ with exactly one endpoint on $c$. 
The replacements must ensure that the total weight of the new edges is no smaller than that of the replaced ones. To describe this process more formally, we define the vertex surrounding and edge surrounding of $c$ (with respect to $C_{\max}$), denoted by $\mathit{sur}^v(c)$ and $\mathit{sur}^e(c)$, respectively.

For a $3$-triangle $t=(p,q,r)$ we have $sur^v(t)=\{p,q,r\}$ and $sur^e(t)=\{(p,q), (q,r), (r,p)\}$. For a $2$-triangle $t=(p;q,r)$, we have $sur^v(t)=\{p,q,r\}$ and $sur^e(t)=\{(q,r), (r,q)\}$.
Let $c$ be a $2$-cycle $(u,v)$ that is a subcycle of a triangle $(u,v,w)$ of $C_{max}$. Then $sur^v(c)=\{w,u,v\}$ and $sur^e(c)=\{(w,u), (u,v), (v,w)\}$.
We say that a multisubgraph $S$ is {\bf \em amenable} on $c$ if:
\begin{enumerate}
\item $S$ has the form $(sur^v(c), E_S)$;
\item $w(E_S) \geq 10w(c)+4w(sur^e(c))$;
\item $S$ is path-$20$-colorable;
\item for any vertex $v \in sur^v(c)$ it holds that $indeg(v) \leq 17$ or $outdeg(v)\leq 17$ in $G_1$ after the replacement of all copies of $c$ and $sur^e(c)$ with $E_S$.
\end{enumerate}

The degrees are required to satisfy the last condition because we want to leave the possibility of adding to the multigraph $G_1$ up to $3$ copies of an edge of $G$ incident to any vertex $v$.
We say that a  cycle $c$ is {\bf \em tricky} if there exists  no  amenable subgraph $S$ on $c$.

\begin{lemma}\label{basiccol}
Let $t=(p,q,r)$ be a triangle consisting of edges $a=(p,q), b=(q,r)$ and $c=(r,q)$ and  let $G_t$ be a multigraph consisting of $x$ copies of $a$, $y$ copies of $b$ and $z$ copies of $c$.  If no $x,y,z$ is greater than $20$ and $x+y+z \leq 40$, then $G_t$ can be path-$20$-colored.
\end{lemma}
\dowod For each edge, we choose which  colors of $\Kd$ are assigned to it. Equivalently, we choose which colors of $\Kd$ do {\emph not} appear on it. Thus $20 -x$ colors will not appear on $a$, $20-y$ will not appear on $b$ and $20-z$ will not appear on $c$. To ensure that $G_t$ is path-$20$-colored it suffices, if each color $k$ of $\Kd$ does not appear on at least one edge of $t$. Since $x+y+z \leq 40$, we get that $60-(x+y+z) \geq 20$. Hence we are able to assign each color $k$ to at least one edge of $t$, where it does not appear.
 \koniec

\begin{fact}\label{rozk}
Let $x$ and $y$ be two nonnegative real numbers such that $x+y=W$ and $x \geq y$ and let $i,j$ be two natural numbers such that $i \geq j$.
Then $ix+jy \geq \frac{i+j}{2}W$.
\end{fact}

\begin{lemma} \label{est}
Let $t$ be a tricky $3$-triangle and $a$ any of its edges. Then $\frac{1}{4}w(t) < w(a) <\frac{2}{5} w(t)$.
Let $p$ be any two-edge path contained in $opp(t)$. Then  $w(p) < \frac{4}{5}w(t)$.
\end{lemma}
\dowod
We show that if at least one of these statements does not hold, then there exists an amenable
subgraph on $t$ (by definition of weight at least $14w(t)$), which contradicts the fact that $t$ is  tricky.

Suppose  that $w(a) = \frac{2}{5} w(t) +\epsilon$ for some edge $a$ of $t$ and $\epsilon \geq 0$. Then the other two edges have total weight $\frac{3}{5} w(t) -\epsilon$. By taking $20$ copies of $a$ and $10$ copies of each of the remaining two edges, by Lemma \ref{basiccol} we obtain a path-$20$-colorable multigraph. Its weight is equal to $20(\frac{2}{5} w(t) +\epsilon)+10(\frac{3}{5} w(t) -\epsilon) \geq 14 w(t)$. The indegree or outdegree of each vertex of $t$ is not greater than $10<17$. Hence, we get an amenable subgraph on $t$.

Suppose now that $\frac{1}{4}w(t) \geq w(a)$ for some edge $a$ of $t$. Then $w(a)=\frac{1}{4}w(t)-\epsilon$ for some $\epsilon\geq 0$ and the other two edges have total weight  $\frac{3}{4}w(t)+\epsilon$.  By taking $8$ copies of $a$, and $16$ copies of each of $b,c$, we obtain an amenable
subgraph on $t$, because its weight  is equal to  $16(\frac{3}{4}w(t)+\epsilon)+8(\frac{1}{4}w(t)-\epsilon)=\frac{48+8}{4}w(t)+8\epsilon$, which is at least $14w(t)$. 

Assume now that there exists a two-edge path  $p$ contained in $opp(t)$ having weight  $w(p) \geq \frac{4}{5}w(t)$. Suppose that this path consists of $a$ and $b$ and that $w(a) \geq w(b)$. Then, by taking $20$ copies of  $a$ and  $15$ of $b$, by Fact \ref{est} we obtain a
subgraph on $t$ of weight at least $(20+15)\frac{2}{5}w(t) \geq 14w(t)$. It is easy to see that this subgraph satisfies also the other two conditions of an amenable subgraph. \koniec

The corollary of the second part of this lemma is the following:

\begin{lemma} \label{est1}
Let $t$ be a tricky $3$-triangle and $a$ any of its edges. Let $a', b',c'$ denote edges of $opp(t)$ such that $w(a')\geq w(b') \geq w(c')$.
Then $w(b') < \frac{2}{5}w(t)$. 
\end{lemma}

\begin{lemma} \label{nontricky}

Let $t=(p;q,r)$ be a tricky $2$-triangle of $C_1$ and let $a=(q,r), d=(r,q), b_1=(p,q), b_2=(r,p)$. Then 

\begin{enumerate}
\item $w(r,q)> \max\{\frac{w(t)}{2}, \frac{3}{2}w(q,r)\}$.
\item $w(r,q)> \frac{w(t)}{2}$.

\item \label{resclem} $\min\{w(p,q), w(r,p)\} > \frac{w(d)}{3}$.

\item $w(d)> \frac{w(b_1)+w(b_2)+w(t)}{2} - w(a)$.
\item $6w(d)+w(a)> 5(w(b_1)+w(b_2))$.
\end{enumerate}
\end{lemma}
\dowod 

If points $1$ and $2$ did not hold, we could replace $4$ copies of $(r,q)$ with two copies of $t$ (and hence $6$ edges of $t$) and obtain a path-$20$-colorable subgraph or replace $4$-copies of $(r,q)$  with $6$ copies of $(q,r)$ and again obtain a path-$20$-colorable subgraph.

We now prove  point $3$. We notice that in order for
$t$ to be tricky it has to hold that $4w(c)+10w(t) > 10w(a)+10w(b_2)+10w(d)$,  because the subgraph consisting of $10$ copies of each of $a, b_2,
d$ is path-$20$-colorable. This is because we can use $10$ colors on $d$ and the remaining ones on $a$ and $b_2$. This means that $10 w(b_2) > 6w(d)-4 w(a)$, which implies that $w(b_2) > \frac{3}{5}w(d) -\frac{2}{5}w(a)$.
Since $w(d)> \frac{3}{2}w(a)$, we obtain that $w(b_2)>\frac{w(d)}{3}$ and $w(b_2) > \frac{w(a)}{2}$.
We obtain the same estimation for $w(b_1)$.


To prove point $4$, we consider a subgraph consisting of $14$ copies of each of $b_1$ and $b_2$ and $12$ copies of $a$.
If this subgraph is not amenable, then $14w(b_1)+14w(b_2)+12w(a) < 10w(t)+4w(a)+4w(d)$, which implies that $4w(d) > 2(w(b_1)+w(b_2))+2 w(t) - 4 w(a)$,
which is equivalent to $w(d)> \frac{w(b_1)+w(b_2)+w(t)}{2} - w(a)$.

The last point follows from combining point $4$: $4w(d) > 2(w(b_1)+w(b_2))+2 w(t) - 4 w(a)$ with point $1$: $w(d) > \frac{3}{2}w(a)$.

\koniec

\section{Construction of $G_1$ in the presence of tricky triangles and tricky $2$-cycles} \label{multi}


We show how to modify the multigraph $G_1$ built in the previous section, when $G$ contains  tricky triangles or $C_1$ contains strange $2$-cycles. 

\begin{figure}[h]
\centering{\includegraphics[scale=0.6]{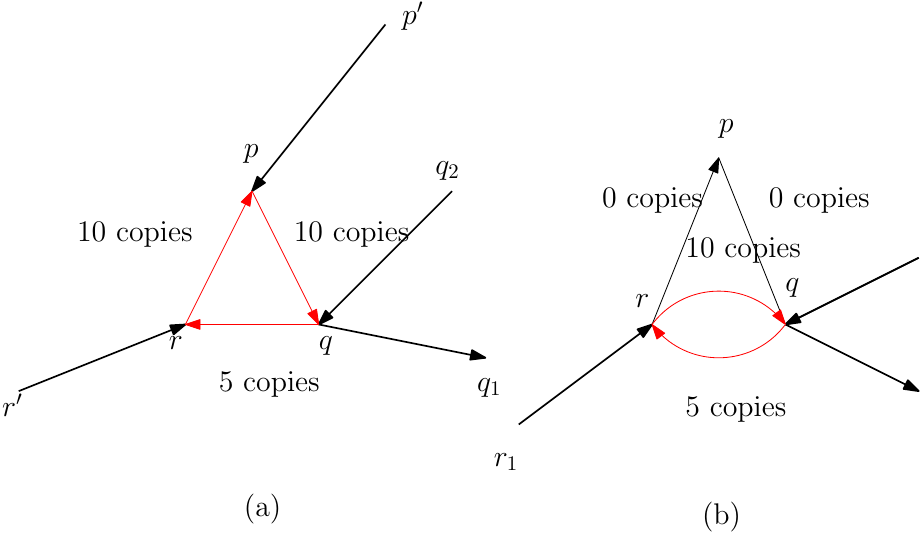}}
\caption{{\scriptsize  Halfy triangles with $3$ incoming and $1$ outgoing edges of $C_1$: a tricky $3$- and $2$-triangle $t=(p,q,r)$.}
} \label{htrgs}
\end{figure}

\begin{figure}[h]
\centering{\includegraphics[scale=0.6]{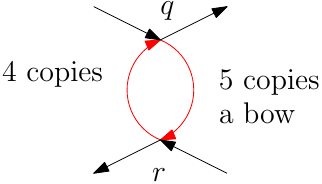}}
\caption{{\scriptsize  A bow.}
} \label{bow}
\end{figure}

The main new features are going to be the following $3$ types of subgraphs, shown in Figures \ref{htrgs}  and \ref{bow}, arising on tricky triangles:
\begin{enumerate}
\item a subgraph on $p,q,r$ such that  $C_{max}$ contains a halfy $3$-triangle $t=(p,q,r)$ and  $C_1$ contains exactly four edges incident to $t$: either three incoming edges and one outgoing of $t$ or three outgoing and one incoming. W.l.o.g. assume that $C_1$ contains
edges $(p',p), (r',r), (q,q_1), (q_2, q)$. We  proceed as indicated in the definition of a harmonious triangle. $G_1$ then contains either (i) $10$ copies of each of $(r,p),(q,r)$ and $5$ copies of $(p,q)$ or (ii) $10$ copies of each of $(p,r),(r,p)$ and $0$ copies of $(p,q), (q,r)$. We choose the option of maximum weight. If option (ii) is maximum, then we treat $C_{max}$ as though it contained the $2$-cycle $(p,r)$ and not the triangle $(p,q,r)$ and in coloring $G_1$ we do not treat $(p,q,r)$ as a tricky triangle.
Any edge $e \in C_{max}$ such that $mult(e)=10$ is called a {\bf \em b-edge}. A subgraph of $G_1$ on $p,q,r$ contains thus two b-edges.

\item a subgraph on $p, q,r$ such that  $C_{max}$ contains a halfy $2$-triangle $t=(p,q,r)$ with a t-cycle $c=(q,r)$ and  $C_1$ contains  exactly four edges incident to $t$, three of which are incident to $c$.  W.l.o.g assume that two edges of $C_1$  are incident to $q$ and  that $C_1$ contains an edge $(r_1, r)$. Then $mult(r,q)=10$ and $mult(q,r)=5$, hence $(r,q)$ is a b-edge. 

\item a subgraph on $q,r$ such that  $C_{max}$ contains a $2$-cycle $c=(q,r)$, which is a t-cycle of a tricky $2$-triangle $t$ and $C_1$ contains a loop $e'_t$. Then $C_1$ contains four edges incident to $c$ and
 $mult(q,r)=5, mult(r,q)=4$. We call the edge $(q,r)$ a  {\bf \em bow}. 

\end{enumerate}

\begin{figure}[h]
\centering{\includegraphics[scale=0.6]{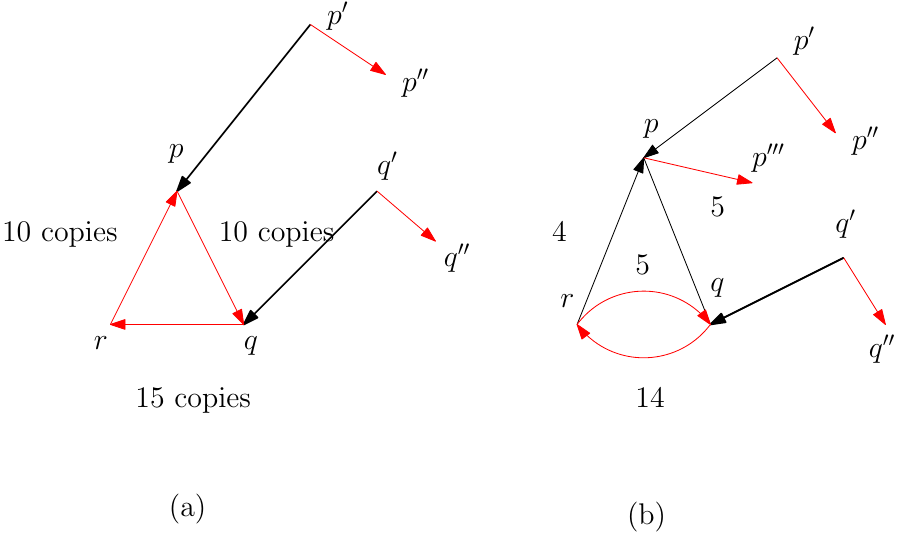}}
\caption{{\scriptsize  Halfy triangles with two incoming edges of $C_1$: a tricky $3$- and $2$-triangle $t=(p,q,r)$.}
} \label{htrgs2}
\end{figure}

The following two types of subgraphs can be treated in a very similar way as a subgraph surrounding a halfy $2$-cycle (Figure \ref{htrgs2}):
\begin{enumerate}
\item a subgraph on $p,q,r$ such that  $C_{max}$ contains a halfy $3$-triangle $t=(p,q,r)$ and  $C_1$ contains exactly two edges incident to $t$: either two incoming or two outgoing. W.l.o.g. assume that $C_1$ contains
edges  $(p',p)$ and $(q',q)$. We again proceed as indicated in the definition of a harmonious triangle. $G_1$ contains then either (i)  $10$ copies of each of $(p,q), (r,p)$ and $15$ copies 
of $(q,r)$ or (ii) $10$ copies of each of $(p,r), (r,q), (q,r)$ or (iii) $10$ copies of each of $(q,r), (r,p), (p,r)$.  We choose the option with maximum weight. If option (ii) or (iii) is maximum, then we treat $C_{max}$ as though it contained the $2$-cycle $(r,q)$ or $(r,p)$ and not the triangle $(p,q,r)$ and in coloring $G_1$ we do not treat $(p,q,r)$ as a tricky triangle. We call the edges $(p', p''), (q', q'')$ of  $C_{max}$ {\bf \em antennas} of $t$ and require that they are diverse.
\item a subgraph on $p,q,r$ such that  $G$ contains a halfy $2$-triangle $t=(p,q,r)$, where $c=(q,r)$ is its t-cycle and  $C_1$ 
contains exactly two edges incident to $t$: either two incoming or two outgoing. W.l.o.g. assume that
 $C_1$ contains
edges  $(p',p)$ and $(q',q)$.  $G_1$ contains then  $5$ copies of each of $(p,q), (r,q)$, $4$ copies of $(r,p)$ and $14$ copies 
of $(q,r)$. We call the edges $(p', p''), (q', q'')$ of  $C_{max}$ {\bf \em antennas} of $t$ and require that they are diverse.
We call the edge $(p,p''')$ of $C_{max}$ a {\bf \em weak antenna} of $t$ and require that it is {\bf \em weakly diverse} with $(p', p'')$, by which we mean that $|col(p, p''') \setminus col(p',p'')| \geq 2$.
\end{enumerate}

If $C_1$ contains some strange $2$-cycle or  tricky triangles, then the multigraph $G_1$  contains non-path-$20$-colorable subgraphs.
We deal with such non-colorable subgraphs at the end by finding exchange sets $F_1, F_2$ and extending the partial path-$20$-coloring.

If $C_1$ contains a $2$-cycle or triangle of $C_{max}$, then such a cycle is not  tricky and then we can replace it
with other edges so as to receive an amenable subgraph.

In all other aspects the obtained multigraph $G_1$ has almost the same properties as in previous sections, i.e., each vertex has at most two incoming and two outgoing edges: two in $C_{max}$ and two in $C_1$  and thus indegree and outdegree at most $14$. 

Below we give a detailed description of the construction of $G_1$.

\subsection{Tricky $2$-cycles}

 Let $c=(u,v)$ be a $2$-cycle which is a subcycle of a triangle $(u,v,x)$ of $C_{max}$.
 The cycle $c$ is tricky if both (i) $w(u,v) > \frac 14 (w(x,u)+w(v,x))$ and (ii) $w(v,u) > \frac 34(w(x,u)+w(v,x))$.

If $C_1$ contains exactly one half-edge of each edge of a $2$-cycle $c$ of $G$, then $c$ is called a {\bf \em halfy} $2$-cycle of $C_1$.
In the way shown below we  deal with  each  halfy $2$-cycle that is a subcycle of a triangle $t$ of $C_{max}$ but $t$ is not tricky.  
To facilitate the subsequent coloring of $G_1$, we modify it as follows. We are also going to modify $C_{max}$. To avoid confusion, we denote the modified $C_{max}$ as $C'_{max}$.  Let $c=(u,v)$ be any such halfy 2-cycle of $C_1$. Suppose that $(u,v) \in C_{max}$. Let $t=(u,v,x) \in C_{max}$. 
We remove all copies of $(x,u), (v,x)$ ($8$ in total) from $G_1$ and replace them with $1$ additional copy of $(u,v)$ and $4$ additional copies of $(v,u)$ - as a result $mult(u,v)=mult(v,u)=10$ and $mult(x,u)=mult(v,x)=0$. Thus, $(u,v)$ and $(v,u)$ are b-edges. Also, in $C'_{max}$, we replace edges $(x,u), (v,x)$ with one edge $(v,u)$. As a consequence of this modification, we obtain a multigraph $G_1$, in which each halfy $2$-cycle $c=(u,v)$ of $C_1$ has no incident edges of $C_{max}$ apart from those already belonging to $c$  and both edges of $c$ are b-edges. 

\begin{lemma}
The weight of the thus modified $G_1$ is at least $4w(C_{max})+10w(C_1)$.
\end{lemma}
\dowod This follows from the fact that each halfy  $2$-cycle $(u,v)$ of $C_1$ is tricky.  Assuming that it is $(u,v)$ that belongs to $C_{max}$ by the definition we get that $w(u,v) > \frac 14 (w(x,u)+w(v,x))$ and  $w(v,u) > \frac 34(w(x,u)+w(v,x))$, which means that $w(u,v)+5w(v,u)\geq 4(w(x,u)+ w(v,x))$.

\koniec

\subsection{Tricky $3$-triangles}

For any tricky $3$-triangle, which is halfy in $C_1$, we proceed as described at the beginning of this section. 
This is justified by Lemma \ref{relax}, which says that any such triangle is harmonious in $C_1$.

\subsection{Tricky $2$-triangles}

Let $t=(p,q,r)$ be a tricky triangle  such that $c=(r,q)$ is a $2$-cycle of $C_{max}$.
Let $a=(q,r), d=(r,q), b_1=(p,q), b_2=(r,p)$.

\begin{enumerate}
\item $C_1$ contains  two crossing half-edges within $b_1, b_2$ and no half-edges within $c$.

Then $C_1$ contains either $(r_1,r)$ and two edges incident to $q$ or $(q_1,q)$ and two edges incident to $r$. W.l.o.g. assume that the first case holds.
$G_1$ contains then $10$ copies of $d$ and $5$ copies  of $a$. 
Suppose to the contrary that $10w(d)+5w(a) < 5w(b_1)+5w(b_2) + 4w(c)$. Then $10w(d)+5w(a) < 5w(b_1)+5w(b_2) + 4w(a)+4w(d)$, which implies that $6w(d)+w(a)<5w(b_1)+5w(b_2)$. Hence, we  contradict  Lemma \ref{nontricky} point $5$. Therefore, 
$10w(d)+5w(a) \geq 5w(b_1)+5w(b_2) + 4w(c)$.

\item $C_1$ contains  two crossing half-edges within $b_1, b_2$ and two crossing half-edges within $c$.

Then $C_1$ contains either (i) $(p_1,p)$ and $(q,q_1)$ or (ii) $(r_1,r)$ and $(p,p_1)$. W.l.o.g. assume that
case (i) holds. Then $G_1$ contains $15$ copies of $d$ and depending on which is more convenient either (i)
$5$ copies of $a$ and $5$ copies of that edge from $b_1, b_2$ which has greater weight or (ii) $5$ copies of each of 
$b_1, b_2$.  To facilitate the coloring of $G_1$, we modify $G'_1$ in such a way that we replace edges $(p,q), (q,q_1)$ with one edge
$(p,q_1)$. The edge $(p,q_1)$ has multiplicity $10$ in $G'_1$. $G'_1$ does not contain any of the remaining edges of $t$ or $c$. The restriction regarding this modification is such that if $G'_1$ contains $(q_1,p)$ and $mult(q_1,p)=14$, then we do not perform  it and instead remove from $G'_1$ all edges of $t$ and $c$ as well as $(q,q_1)$.

Note that the required weight is equal to $W=4w(c)+5w(c)+5w(b_1)+5w(b_2)=9w(a)+9w(d)+5w(b_1)+5w(b_2)$.
We need to prove that $15w(d)+5w(a)+5\max\{w(b_1), w(b_2)\} \geq W$ and $15w(d)+5w(b_1) + 5w(b_2) \geq W$.

Let us prove the first one. Notice that $\max\{w(b_1), w(b_2)\} \geq \frac{w(t)-w(a)}{2}$. Suppose that $15w(d)+5w(a)+5\max\{w(b_1), w(b_2)\} < W$. This means that $4w(a)+ 5\min\{w(b_1), w(b_2)\} > 6w(d)$.  Using $w(d)=\frac{w(t)}{2}+\epsilon$, we get that 
$1.5 w(a)+2.5w(t) > 3w(t)+6\epsilon$. Hence $w(a)>\frac{w(t)}{3}+4\epsilon$. But then $w(d) < \frac{3}{2}w(a)$, which contradicts Lemma \ref{nontricky} point $1$.

To prove the second one suppose that $15w(d)+5w(b_1) + 5w(b_2) < W=9w(a)+9w(d)+5w(b_1)+5w(b_2)$. This is equivalent to $6w(d)< 9w(a)$, which again contradicts Lemma \ref{nontricky} point $1$.

\item $C_1$ contains two crossing half-edges within $(p,q), (r,p)$ and one whole edge within $c$.

Then $C_1$ contains either (i) $(p_1,p)$ and $(q_1,q)$ or (ii) $(p,p_1)$ and $(q,q_1)$. The case is similar to the case  for the tricky
$3$-triangle. We do not modify anything in $G_1$.
 W.l.o.g. assume that case (i) holds.  Edges $(p_1, p_2), (q_1, q_2)$ of $C'_{max}$ are called the {\bf \em antennas} of $t$ and required to be diverse.

\item $C_1$ does not contain any half-edges within $c$ but contains a loop $e'_t$. 

Since $10w(e'_t)=w(r,q)$, $G_1$ contains $5$ copies of $(r,q)$ (and not $4$ as usual). Such an edge $(r,q)$ is called a {\bf \em bow}.

\item $C_1$ contains all edges of $t$ and a loop $e_t$.

Then $10w(t)+4w(c)+10w(e_t)= 14w(q,r)+10(w(p,q)+w(r,p))+3w(r,q)$ and $G_1$ indeed contains $14$ copies of $(q,r)$, $10$ copies of each of
$(p,q), (r,p)$ and $3$ copies of $(r,q)$, which is a path-$20$-colorable subgraph.
\end{enumerate}

\begin{theorem}
$w(G_1) \geq 10w(C_1)+4w(C_{max})$
\end{theorem}
The proof follows from the above discussion.

\section{Exchange sets and completing the path-coloring} \label{completing}
\subsection{$2$-cycles of $C_1$ sharing an edge with $C_{max}$}

We are going to use the following crucial lemma.

\begin{lemma} \label{altpath}
Let $P$ be an alternating path $(f_1=(u_1, v_1), e_1, f_2, \dots, e_k, f_{k+1}=(u_{k+1}, v_{k+1}))$ of $C_{\max} \oplus C_1$ with edges $f_1, f_2, \dots, f_{k+1}$ belonging to $C_{\max}$. Then:
\[
\sum_{i=1}^{k+1} w(f_i) \ge \sum_{i=1}^{k} w(e_i).
\]
\end{lemma}

\dowod
Suppose to the contrary that $\sum_{i=1}^{k+1} w(f_i) < \sum_{i=1}^{k} w(e_i)$. We will show that in this case, we can construct a valid cycle cover $C$ with a weight strictly greater than $w(C_{\max})$.

Let us add the edge $(u_{k+1}, v_1)$ to the path $P$, thereby forming an alternating cycle $P'$. Since all edges in the graph have nonnegative weights, it follows that:
\[
\sum_{i=1}^{k+1} w(f_i) \le \sum_{i=1}^{k} w(e_i) + w(u_{k+1}, v_1) = w(P' \cap C_1).
\]
Hence, the modified cycle cover $C = C_{\max} \oplus P'$ satisfies $w(C) > w(C_{\max})$, yielding a contradiction to the maximality of $C_{\max}$.

Finally, note that in the case where $u_1 = v_{k+1}$, the vertex $u_1$ has no incident edges in $C$ and thus  forms an isolated vertex (a cycle of length $0$).
\koniec

Let $S$ denote the set of $2$-cycles of $C_1$ that share an edge with $C_{\max}$ but are not contained in any problematic cycle. 
We now define so-called \emph{\textbf{exchange sets}} $E_1$ and $F_1$ and a function $f: E_1 \rightarrow F_1$. Let $c = (u,v)$ be any $2$-cycle of $S$ such that $(u,v) \in C_{\max}$. Then $E_1$ contains exactly one edge—namely, the edge $(v,u) \in c \setminus C_{\max}$—and its image $f((v,u))$ consists of two edges, $(u', u)$ and $(v, v')$, belonging to $C_{\max}$. We define $F_1$ as $f(E_1)$.

\begin{lemma}
The exchange sets $E_1, F_1 \subseteq E$ satisfy the following condition:
\[
w(E_1) \le w(F_1).
\]
\end{lemma}
\dowod 
The edges of $E_1 \cup F_1$ can be decomposed into a collection of alternating paths and cycles in $C_{\max} \oplus C_1$. Therefore, applying Lemma~\ref{altpath} and Fact~\ref{altfact}, it directly follows that $w(E_1) \le w(F_1)$.
\koniec

Let $G^1_1$ denote the multigraph $G_1$, in which we remove $4$ copies of each edge of $E_1$ and add $4$ copies of each edge of $F_1$.

\begin{figure}[h]
\centering{\includegraphics[scale=0.75]{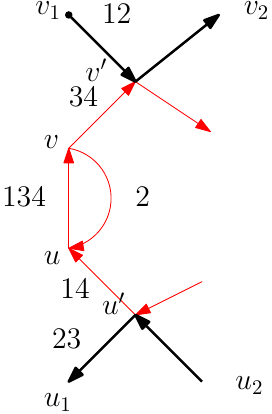}}
\caption{Extension of path-$4$-coloring}
 \label{2cycleG11}
\end{figure}

\begin{lemma}
The path-$4$-coloring of the multigraph $G_1$ can be extended to the path-$20$-coloring of $G^1_1$.
\end{lemma}
\dowod
Before coloring $G_1$ we modify it as follows.
Let $P=(u', u, v, v')$ be a path consisting of edges of $C_{max}$ such that  the edge $(v,u)$ belongs to $E_1$ and the edges $(u',u), (v,v')$ belong to $F_1$. We replace $P$
with one edge $e_P=(u',v')$.
(We proceed analogously if a path is longer and its every other edge belongs to $E_1$. More precisely, let $P=(u_1, u_2, u_3, \ldots, u_k, u_{k+1})$ be a path consisting of edges of $C_{max}$ such that (i) $(u_2, u_3), (u_4, u_5), \ldots, (u_{k-1}, u_{k})$ belong to $E_1$, (ii) $(u_1, u_2)$ belongs to the image of only one edge of $E_1$ - $(u_2, u_3)$, (iii)  $(u_k, u_{k+1})$ belongs to the image of only one edge of $E_1$ - $(u_{k-1}, u_{k})$.  We then replace $P$
with one edge $e_P=(u_1, u_{k+1})$.)
In the path-$4$-coloring of the modified multigraph $G_1$ the edge $e_P$ will be colored with one color of $\Kr$.  Let $(u',u_1), (v_{1}, v')$ denote two edges of $C_1$. Each of them is assigned some two colors of $\Kr$. Since $e_P$ is colored with one color,   $col(u', u_1)\cap col(v_{1}, v')$ contains at least one common color.  The extension to the path-$4$-coloring of the original $G_1$ (and next to the path-$20$-coloring of $G^1_1$) is most convenient when 
$|col(u', u_1)\cap col(v_{1}, v')|=1$. This is because of the following. See Figure \ref{2cycleG11}. Suppose that $col(v_1, v')=\{1,2\},  col(u',u_1)=\{2,3\}, col(u',v')=4$. Then, we can assign $4$ to each of the edges $(u',u), (u,v), (v,v')$ of the path $P$.
The second color assigned to $(u',u_1)$ is going to be $1$ and to $(v,v')$ - $3$. We also assign $1$ and $3$ to $(u,v)$ - hence $col(u,v)=\{1,3,4\}$. Finally, we assign $2$ to $(v,u)$. Such path-$4$-coloring can be easily transformed into path-$20$-coloring in the following manner. Let us recall that in $G^1_1$ we have $mult(u',u)=mult(u,v')=8, mult(v,u)=6, mult(u,v)=16$. For each of the colors $1,2,3,4$ of $\Kr$, first we replace it with $5$ colors of $\Kd$ - color $i$ is replaced with $5i, 5i-1, 5i-2, 5i-3, 5i-4$. Next, we remove colors $19,20$ from $(u',u), (v, v')$ and we transfer color $20$ from $(u,v)$ to $(v,u)$. We can easily check  that we have not introduced any monochromatic cycle.

To guarantee that $|col(u', u_1) \cap col(v_{1}, v')|=1$, we either ensure that the edges $(u'_1, u_1), (v_1, v'_1)$ of $C_{max}$ get distinct colors or we modify $G_1$ further by replacing  $(u',u_1)$ and $(u_2, u')$ with one edge $(u_2, u_1)$and by replacing $(v_1, v'), (v', v_2)$ with $(v_1, v_2)$.
\koniec

\subsection{Tricky $2$-triangles}

Let $c=(r,q)$ be a $2$-cycle contained in a tricky $2$-triangle $t=(p,q,r)$
To the $2$-cycle $c$ we assign weight $w'(c)=w(q,r)$ and by  $\kappa(t)$ we denote $\frac{w(r,q)}{10}$. 

Let $R'$ denote the set of all tricky triangles of $C_1$. They correspond to a matching $N'$ of $H$. Notice that $N \cap N' =\emptyset$, because no tricky triangle of $R$ (corresponding to $N$) can occur in $C_1$. Thus $N \cup N'$ forms a set of alternating paths and cycles. Since $N$ is a  maximum matching of $H$, on each alternating path $P$     the  number of edges of $N'$ on $P$ equals or is smaller than the number of edges of $N$. Hence, w.l.o.g. we may assume that each alternating path has even length.
For each alternating cycle and each alternating path of even length we  replace some edges of triangles of $R'$ with edges belonging to triangles represented by edges of $N$ belonging to the same path or cycle. More precisely, suppose that an alternating path $P$ or cycle $C$ consists of a sequence of edges $e_1, f_1, \ldots,
e_i, f_i, \ldots, e_k, f_k$ such that for $1 \leq i \leq k$ it holds that  $e_i \in N', f_i \in N$ and edges $e_i, f_i$ have a common vertex in $V(H)$. Then we replace some edges of  each tricky triangle $t_i$ of $C_1$ corresponding to edge $e_i$ with some edges of a tricky triangle (not occurring in $C_1$) corresponding to edge $f_i$.

We now describe the exact procedure of replacement. 
Let $t_i=(p,q,r)$ be a tricky triangle of $C_1$ with a t-cycle $c_i=(q,r)$. In $G_1$ we take $14$ copies
of $(q,r)$, $10$ copies of each of $(p,q), (r,p)$ and $3$ copies of $(r,q)$. This means that we are lacking only one copy of $(r,q)$, i.e.,:
\begin{fact}
The weight of the induced subgraph $G_1(t_i)$ of $G_1$ on vertices $p,q,r$ satisfies: \\

$w(G_1(t_i)) = 4w(c_i)+10w(t_i) -  w(r,q)$.
\end{fact}

Consider  alternating paths  and cycles of $N \cup N'$. Each one of them consists of some sequence of edges $e_1, f_1, \ldots,
e_i, f_i, \ldots, e_k, f_k$. For any alternating cycle $C$, we can additionally arrange the  edges on  $C$ so that that a common vertex of any two edges $e_i$ and $f_i$ on $C$ in $H$ corresponds to a $2$-cycle $c_i$. Let $(e_i, f_i)$ be any pair of edges  from such alternating cycle or path and suppose that a tricky triangle $t_i$ of $C_1$ corresponding to $e_i$ has  the form $t_i=(p,q,r)$. If the common vertex of $e_i$ and $f_i$ in $H$ corresponds to a $2$-cycle $c_i$, then a tricky triangle represented by $f_i$ has the form $t'_i=(p', q,r)$. We add   either $(p',q)$ or $(r,p')$ to $F_2$ (and also $3$ copies of the edge added to $F_2$ to $G_1$). 
If, on the other hand, the common vertex of $e_i$ and $f_i$ in $H$ corresponds to the vertex $p$, then a tricky triangle represented by $f_i$ has the form $t'_i=(p, q',r')$. In this case we add   either $(p,q')$ or $(r',p)$ to $F_2$ (and also $3$ copies of the chosen edge to $G_1$). 
We call a tricky triangle $t'_i$ of $R$  corresponding to the edge $f_i$ a {\bf \em rescuer} of $t_i$.

By Lemma \ref{nontricky} point $(3)$ we get that the addition of $3$ copies of each edge of $F_2$ compensates for the lacking copies $(r,q)$ of each tricky $2$-triangle $(p;q,r)$.

We show that we are able to extend the current path-coloring of $G_1$ to the subgraphs containing  tricky $2$-triangles of $C_1$. 
We proceed in the order dictated by directed paths and cycles of a graph  $H^{dir}$, which is a compressed
and directed version of the graph $H$. $H^{dir}$ is obtained from $H$ as follows. For each tricky triangle $t$ of $C_1$ we  identify as one vertex $v_t$ three  vertices in total: all vertices of $t$ as well as the t-cycle $c$ of $t$. 
Let $e=(u,v)$ be any edge of $N$. It then corresponds to a tricky triangle $t'$ of $R$. If $t'$ is a rescuer of a tricky triangle $t$ of $C_1$, we direct the counterpart of $e$ in $H^{dir}$ from $v_t$. 

We fist deal with directed cycles of $H^{dir}$.

\begin{figure}[h]
\centering{\includegraphics[scale=0.75]{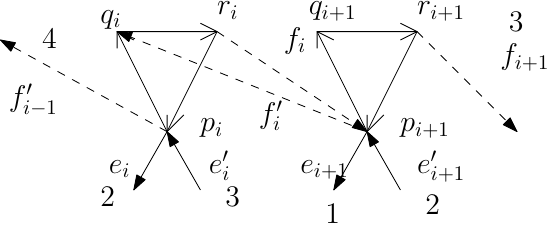}}
\caption{ The edges $e_i, e'_i, e_{i+1}, e'_{i+1}$ are colored correspondingly with $2,3,1,2$, the edge $f'_{i-1}$ is colored with $4$ and the edge $f_{i+1}$ with $3$. Therefore, we color $f'_i$  with $4$. }
 \label{dependcycle}
\end{figure}

\begin{lemma}
Let $c_H$ be any directed cycle of $H^{dir}$. We are able to extend the partial coloring of $G_1$ to the edges of tricky triangles covered by $c_H$ and  the edges of $F_2$ of their rescuers.
\end{lemma}
\dowod Let $t_1, \ldots, t_k$ be the order of tricky triangles of $C_1$, in which they (or more precisely, the vertices representing them) occur on $c_H$. Assume that each $t_i$ has the form $t_i=(q_i, r_i, p_i)$, where $(q_i, r_i)$ is a t-cycle of $t_i$. This means that 
for each $1 \leq i\leq k$ a rescuer $t'_i$ of $t_i$ has the form $t'_i=(q_i, r_i, p'_i)$, where $p'_i$ lies on $t_{i+1}$ (indices are taken modulo $k$). The vertex $p_i$ is incident to two edges $e'_i=(s_i, p_i), e_i=(p_i, s'_i)$ belonging to $C_{max}$, which are already colored. We can assume that they are colored with distinct colors. See Figure \ref{dependcycle}.
We assign $1$ color of $\Kr$ (which translates to $3$ colors of $\Kd$) either to $f_i=(r_i, p'_i)$ or to $f'_i=( p'_i, q_i)$. We do it in such a way that:
\begin{itemize}
\item A color assigned to $f_i$ or $f'_i$ does not occur on any of  $ e_{i+1}, e'_{i+1}, f_{i-1}, f'_{i-1}$.
\item A color occurring on $e_i$ may be assigned to $f'_i$ but not $f_i$. Similarly, a color occurring on $e'_i$ may be assigned to $f_i$ but not $f'_i$.
\end{itemize}

We now show that we are able to assign colors to each $f_i$ or $f'_i$ to satisfy the above. Suppose that we consider $t_i$ for $i<k$.  Assume also that $t_{i-1}$ and $t_{i+1}$ were considered before and thus one of $f_{i-1}, f'_{i-1}$ and one of $f_{i-1}, f'_{i-1}$ are already colored
with $1$ color. Since $t_i$ has $3$ colored edges incident to it and so does $t_{i+1}$ and there are $4$ colors, there exists a color $k$ that does not occur on eny edge incident to $t_{i+1}$ and we assign $k$ either to $f_i$ or to $f'_i$ - in such a way that $k$ can be assigned to two edges of $t_i$. Foe example, in Figure \ref{dependcycle}
we assign $4$ to $f'_i$. 

Once we have assigned colors to rescuers, we can easily color the edges of the triangles and replace one color on $f_i$ or $f'_i$ with some $3$ of its subcolors, i.e.,
$1,2,3,4,5$ are subcolors of $1$, $6,7,8,9,10$ are subcolors of $2$ and so on. Thus, for $t_i$ we have at most $6$ colors from rescuers (one of $t_i$ and one of $t_{i-1}$ and $8$ colors from $e_i, e'_i$. 
\koniec

\begin{lemma}
Let $p_H$ be any directed path of $H^{dir}$. We are able to extend the partial coloring of $G_1$ to the edges of tricky triangles covered by $p_H$ and  the edges of $F_2$ of their rescuers.
\end{lemma}
\dowod

To ensure that for any tricky triangle $t_i$ on a path of $H^{dir}$  it holds that at the moment of coloring $t_i$ at most one colored edge of $F_2$ is coincident with the edges $e_i, e'_i$ and thus to ensure that we have enough colors at our disposal,  we compute the following order of dealing  with paths of $H^{dir}$.
We introduce a directed graph $H^P$, in which the vertex set consists of paths of $H^{dir}$ and there exists an edge $(p_1, p_2)$ only if the last edge of $F_2$ incident to $p_1$ is coincident with an edge $e_i$ or an edge of $e'_i$ such that $e_i, e'_i$ are incident to a triangle on $p_2$. Let us notice that $H^P$ consists 
of vertex-disjoint paths and cycles.

  Let $t$ have the form $t_i=(p_i; q_i, r_i)$. Then
a rescuer $t'$ of $t$ has the form $t'=(q_i, r_i, p'_i)$.  The vertex $p_i$ is incident to two edges $e_i=(s_i, p_i), e'_i=(p_i, s'_i)$ belonging to $C_{max}$, which are already colored. We may assume that $e_i$ and $e'_i$ are colored with distinct colors, say $1$ and $2$. (It is easy to enforce that by replacing $t_i$ with a $2$-cycle of $C_1$.) However, each one of them may also be recolored with one color, say that these are again colors $1$ and $2$. (If these colors are different, the procedure of completing the coloring may only become easier.) It may happen that for at most one of these colors, we have already used some $3$ colors as a rescuer of some different triangle. Suppose that $\tilde{f}_i$ is coincident with $e'_i$ and  is also colored with three subcolors of $1$ or $2$. Let $(p'_i, q_i)$ be an edge that we would like to color with $3$ colors. Suppose that all available colors are subcolors of $1$ (and there are at least $4$ of them).  The three subcolors of $1$ assigned to $(p'_i, q_i)$ will also have to be assigned to $(q_i, r_i)$ and $(r_i, p_i)$. Therefore they cannot occur on $e'_i$ and if any one of them does occur on $e_i$ we want to ensure that we do not create a monochromatic cycle. Suppose that $\tilde{f}_i$ is colored with the subcolors $678$ of $2$ and that the colors available for $(p'_i, q_i)$ are the subcolors $1234$ of $1$. 
Since we need to use $8$ subcolors on $e_i, e'_i$ and there are $5$ subcolors of $2$, we will have to use $3$ subcolors of $1$. Therefore, at least one subcolor of $1$ will appear on two of the edges $e_i, e'_i, (p'_i, q_i)$. We consider first coloring $e_i$ with subcolors $6784$ and $e'_i$ with $9,10,1,5$ and $(p'_i, q_i)$ with $234$. If we do not create a monochromatic cycle of color $4$, we retain this coloring. Otherwise, we color $(p'_i, q_i)$ with $123$ and $e'_i$ with $9,10,4,5$. Since there exists  a monochromatic cycle in the first case, $G_1$ contains a path of subcolor $4$ from $s_i$ to $p'_i$, which means that in the second case we do not create a monochromatic cycle because this path is extended to $e_i$ and $e'_i$ and ends on $p'_i$.

It may happen that $mult(e_i)=mult(s_i, p_i)= 8$, because $e_i$ is part of the exchange set of a $2$-cycle. In such a case we may use the property that $2$ subcolors of $1$ or $2$ are guaranteed to be immune to forming part of  a monochromatic cycle. This is because whenever a rescuer has its t-point in vertex $v$
-see Figure \ref{2cycleG11}, we only allow using the edges incoming to $v$. Hence, the subcolors of one color of $(v,v')$ are immune to forming part of  a monochromatic cycle.
We then proceed analogously as in the case above.

 \koniec

\section{Deferred proofs} \label{miss}

\subsection{Relaxed cycle covers}

{\bf Proof of Lemma \ref{relax}}

First we show that any  perfect matching $M$ of $G'$ yields a quasi relaxed cycle cover. For any edge $(u_{out}, e^1_{uv}) \in M$, we add a half-edge $(u, x_{(uv)})$ to $\tilde C$ and for any edge $(u_{in}, e^2_{vu}) \in M$, we add a half-edge $(x_{(vu)},u)$ to $\tilde C$. Since $M$ is a perfect matching of $G'$ each vertex $u_{in}$ and each vertex $u_{out}$ has an incident edge in $M$. Hence, each vertex in $V$
has exactly one outgoing and one incoming half-edge in $\tilde C$. 

If an edge $(u,v) \in E$ does not belong to any problematic cycle, then it is replaced with  an edge $(u_{out}, v_{in})$ in $E'$. Thus, if 
$\tilde C$ contains only one half-edge of some edge $e$, then $e$ must belong to a problematic cycle.

For every tricky $3$-triangle $t$, $G'$ contains eight additional vertices $\gamma^{t+}_p, \gamma^{t-}_p, \ldots, \gamma^c_r, \gamma^c_q$  each of which  excludes exactly one half-edge within $t \cup opp(t)$ from $\tilde C$. This exclusion follows from the fact that each of these vertices is matched to some vertex  $e^1_{(uv)}$ or  $e^2_{(uv)}$, such that $(u,v)\in t \cup opp(t)$. If one of these additional vertices is matched to a vertex  $e^1_{(uv)}$, then a half-edge $(u, x_{(uv)})$ does not belong to $\tilde C$. Similarly,   if it  is matched to a vertex  $e^2_{(uv)}$, then a half-edge $(x_{(uv)}), v)$ does not belong to $\tilde C$. Therefore, for each tricky $3$-triangle $t$, at least eight half-edges within $t \cup opp(t)$ do not belong to $\tilde C$, which means that $\tilde C$ contains at most four half-edges within $t \cup opp(t)$, hence  contains neither $t$ nor $opp(t)$.

Next, we deal with the properties stated in the current lemma.
The proof of the first property is very similar to the proof of Lemma 2 in \cite{PEZ}.

 We denote the half-edges $(u,x_{(u,v)})$ and $(x_{(u,v)},v)$ as $u^\rightarrow_v$ and $v^\leftarrow_u$.
\begin{claim} \label{claim1} Let $t=(p,q,r)$ be a tricky $3$-triangle such that $(r,q)$ is its diago.
W.l.o.g. we may assume that $M$ yields $\tilde C$ that  is integral on $\{(q,p), (r,p)\}$.
\end{claim}
\dowod
Let us observe that if $\tilde C$ is not integral on $\{(q,p), (r,p)\}$, then it contains two crossing half-edges within  $\{(q,p), (r,p)\}$. 
This is because $\tilde C$ may be non-integral on $\{(q,p), (r,p)\}$ only when (i) $\gamma^{t+}_p$ is matched to $e^1_{rp}$ and $\gamma^{t-}_p$  to $e^2_{qp}$ or
(ii) $\gamma^{t+}_p$ is matched to $e^1_{qp}$ and $\gamma^{t-}_p$  to $e^2_{rp}$. In case (i) $\tilde C$ contains half-edges $(q_{out}, e^1_{qp})$ and $(p_{in}, e^2_{rp})$ and we may replace them in $\tilde C$ with the edge $(q,p)$, obtaining thus a quasi relaxed cycle cover of the same weight - recall that half-edges $(p_{in}, e^2_{rp})$ and $(p_{in}, e^2_{qp})$ have equal weight.
 Similarly, in case (ii) $\tilde C$ contains half-edges $(r_{out}, e^1_{rp})$ and $(p_{in}, e^2_{qp})$ and  we may replace them with the edge $(r,p)$. \koniec

Similar reasoning can be applied to prove that:

\begin{claim}
Let $c$ denote the $2$-cycle $(q,r)$.
If vertices $\gamma^c_q$ and $\gamma^c_r$ are matched in $M$ to the subdivision vertices of the same edge of $c$, then $M$ yields
a relaxed cycle cover $\tilde C$, which is integral on $t$.
\end{claim}
\dowod If vertices $\gamma^c_q$ and $\gamma^c_r$ are matched to subdivision vertices of the edge $(q,r)$, then vertices $\gamma^{t+}_q, \gamma^{t-}_q$ must be matched to subdivision vertices of the edge $(p,q)$. If $\gamma^{t+}_r, \gamma^{t-}_r$ are not matched to the subdivision vertices of the same edge, then $M$ yields a quasi relaxed cycle cover $\tilde C$ with two crossing half-edges within $\{(q,r), (p,r)\}$ and we may replace them with one edge in the same manner as in the proof of Claim \ref{claim1}. 

 If, on the other hand, vertices $\gamma^c_q$ and $\gamma^c_r$ are matched to subdivision vertices of the edge $(r,q)$, then vertices $\gamma^{t+}_r, \gamma^{t-}_r$ must be matched to subdivision vertices of the edge $(p,r)$. If $\gamma^{t+}_q, \gamma^{t-}_q$ are not matched to the subdivision vertices of the same edge, then we get a quasi relaxed cycle cover $\tilde C$ with two crossing half-edges within $\{(p,q), (r,q)\}$ and we may again replace them with one edge. \koniec

We now examine which configurations of half-edges within $t \cup opp(t)$ can occur in $\tilde C$ when vertices $\gamma^c_q$ and $\gamma^c_r$ are not matched in $M$ to the subdivision vertices of the same edge of $c$.

Assume first that $\gamma^c_q$ and $\gamma^c_r$ are matched in $M$ to $e^2_{qr}$ and $e^2_{rq}$. This means that $\gamma^{t-}_q$ must be matched to
$e^2_{pq}$ and $\gamma^{t-}_r$ to $e^2_{pr}$. Vertices $\gamma^{t+}_q, \gamma^{t+}_r$ may be matched in $M$ in one of the following four ways:
\begin{enumerate}
\item  $e^1_{pq}$ and $e^1_{pr}$. Thus, $e^1_{qr}$ and $e^1_{rq}$ must be matched to $q_{out}$ and $r_{out}$, which in turn means
that neither can $e^1_{qp}$  be matched to $q_{out}$ nor $e^1_{r,p}$ to $r_{out}$.
 Therefore, $\tilde C$ contains two outgoing half-edges within the $2$-cycle $(q,r)$ and  no half-edges within $\{(p,q), (p,r), (q,p), (r,p)\}$.
\item  $e^1_{rq}$ and $e^1_{qr}$. This would mean that both $e^1_{pq}$ and $e^1_{pr}$ must be matched to $p_{out}$, which is impossible.
Therefore this case  cannot occur.
\item $e^1_{rq}$ and $e^1_{pr}$. 
$\tilde C$ contains half-edges $ (q_{out}, e^1_{qr}),  (p_{out}, e^1_{pq}), $. In this case $\tilde C$ may additionally contain the edge $(r,p)$.
\item $e^1_{pq}$ and $e^1_{qr}$. $\tilde C$ contains half-edges $(p_{out}, e^1_{pr}),  (r_{out}, e^1_{rq})$.  In this case $\tilde C$ may additionally contain the edge $(q,p)$.

\end{enumerate}

Assume next that $\gamma^c_q$ and $\gamma^c_r$ are matched in $M$ to $e^1_{qr}$ and $e^1_{rq}$. This means that $\gamma^{t+}_q$ must be matched to
$e^1_{pq}$ and $\gamma^{t+}_r$ to $e^1_{pr}$. Vertices $\gamma^{t-}_q, \gamma^{t-}_r$ may be matched in $M$ in one of the following ways:
\begin{enumerate}
\item  $e^2_{pq}$ and $e^2_{pr}$. $\tilde C$ contains two incoming half-edges within the $2$-cycle $(q,r)$. 
\item  $e^2_{rq}$ and $e^2_{qr}$. This means that $e^2_{pq}$ must be matched to $q_{in}$ and $e^2_{pr}$ to $r_{in}$. Therefore, $\tilde C$
contains $(q_{in}, e^2_{pq}),  (r_{in}, e^2_{pr})$ and no half-edge within the $2$-cycle $(q,r)$.
\item $e^2_{rq}$ and $e^2_{pr}$. 
$\tilde C$ contains half-edges $(q_{in}, e^2_{pq}),  (r_{in}, e^2_{qr})$.
\item $e^2_{pq}$ and $e^2_{qr}$. $\tilde C$ contains half-edges $(q_{in}, e^2_{rq}),  (r_{in}, e^2_{pr})$.

\end{enumerate}

In all the above four cases $\tilde C$ may contain additionally one of the edges $(q,p), (r,p)$. \\

Next we want to show that $10w(\tilde C)_t +4w(t)$ is upper bounded
by: 
\begin{itemize}
\item 
 $\max\{10w(t) -5w(r,p), 
10(w(p,q)+w(q,p))\}$, if $\tilde C$ contains one edge ougoing of $t$, incident to $r$ and three edges incoming to $t$,
\item $ \max\{10w(t) -5w(q,r), 
10(w(p,q)+w(q,p))\}$, if $\tilde C$ contains one edge incoming to $t$ incident to $r$ and three edges outgoing of $t$, 

\item 
$\max\{10w(t) +5w(r,p), 10( w(p,r)+w(p,q)+w(q,p)), 10(w(p,q)+w(p,r)+w(r,p))\} $, if $\tilde C$ contains two edges outgoing of $t$, incident to $p$ and $q$ and no edge incoming  to $t$,

\item $\max\{10w(t) +5w(q,r), 10( w(q,r)+w(r,p)+w(p,r)), 10( w(p,r)+w(r,q)+w(q,r))\}$, if $\tilde C$ contains two edges incoming to $t$ incident to $p$ and $q$ and no edge outgoing of $t$. 

\end{itemize}

Suppose first that $\tilde C$ is not integral and contains two half-edges within $t \cup opp(t)$. By Claim \ref{claim1} none of these half-edges belongs to
$(q,p)$ or $(r,p)$. If $\tilde C$ does not contain a half-edge of $(r,q)$ (the heaviest edge of $opp(t)$), then to prove that $t$ satisfies condition (iii) it suffices to show that for any two different edges $a',c'$ of $\{(p,q), (p,r), (q,r)\}$ and for any edge $a$ of $t$ it holda that $5w(a')+5w(c')+4w(t) \leq 10w(t)-5w(a)$. This is equivalent to showing $5w(a')+5w(c')+5w(a) \leq 6w(t)$.

\begin{claim}\label{wagixyz}
Let $z$ be any edge of $t$ and let $S =\{x,y\}$ be a set consisting of  two edges of $t \cup opp(t)$ satisfying the condition: if the heaviest edge of $opp(t)$ belongs to $S$, then the other edge of $S$ also belongs to $opp(t)$.  Then $w(x)+w(y)+w(z) \leq \frac{6}{5}w(t)$.
\end{claim}
\dowod
By Lemma \ref{est} we have the following: (i)
 if $x,y \in opp(t)$, then $w(x)+w(y) <\frac{2}{5}w(t)$ (ii) $w(z)< \frac{2}{5}w(t)$ and (iii) for any edge $y \in opp(t)$  different from the heaviest edge of $opp(t)$,  $w(y)< \frac{2}{5}w(t)$. This proves the claim. \koniec

Assume now that $\tilde C$ contains two half-edges, one of which is a half-edge of $(r,q)$ and the other is within $t$. It is possible only when $\tilde C$  contains 
one half-edge of each of $(r,q), (q,r)$. Now we prove that for any edge $a\neq (q,r)$ of $t$  it holds that $5w(r,q)+5w(q,r)+4w(t) \leq \max\{10w(t)-5w(a), 10(w(r,q)+w(q,r)\}$.  
Let  $b=(q,r)$ and $b'=(r,q)$. Suppose that $5w(b)+5w(b')+4w(t)>10w(t)-5w(a)$.
This means that $w(b')+w(b)> \frac{6}{5}w(t)-w(a)$. If  $5w(b)+5w(b')+4w(t)$ is also greater than $10(w(b)+w(b'))$, then $w(b)+w(b')< \frac{4}{5}w(t)$. However, these two inequalities imply that $w(a)> \frac{2}{5}w(t)$, which contradicts Lemma \ref{est}.

Let us now consider the cases when $\tilde C$ is not integral and contains four half-edges within $t \cup opp(t)$. Hence, $\tilde C$
contains the whole edge $(r,p)$ or the whole edge $(q,p)$, because $\tilde C$ can contain at most two half-edges within the remaining edges
of $t \cup opp(t)$. We can notice that if $\tilde C$ contains $(r,p)$, then we have an identical  proof as above for the case of two half-edges within $t \cup opp(t)$ because we then add $10w(a)$ to both sides of the inequality. 

\begin{claim}
Suppose that $\tilde C$ is not integral, contains $c'=(q,p)$ and two half-edges within edges $x$ and $y$. Then $10w(\tilde C)_t +4w(t)$ is upper bounded by $10w(t)+5w(b)$.
\end{claim}
\dowod
We want to show that $5w(x)+5w(y)+10w(c') +4w(t)\leq 10w(t)+5w(b)$, which is equivalent to proving that $w(x)+w(y)+2w(c')\leq \frac{6}{5}w(t)+w(b)$.

We analyze now the four cases when $\tilde C$ contains $(q,p)$ and is not integral.
\begin{enumerate}

\item $\tilde C$  contains also two incoming half-edges within the $2$-cycle $(q,r)$, i.e., $x=b$ and $y=b'$. We want to show that $w(b)+w(b')+2w(c')\leq \frac{6}{5}w(t)+w(b)$,
i.e., $w(b')+2w(c')  \leq \frac{6}{5}w(t)$. We easily obtain it from Lemma \ref{est}, because $w(b')+w(c')< \frac{4}{5}w(t)$ and $w(c') < \frac{2}{5}w(t)$.

\item  $\tilde C$ contains $(q_{in}, e^2_{pq}),  (r_{in}, e^2_{pr})$, i.e. $x=a'$ and $y=c$. Then we show that $w(c)+w(a')+2w(c') \leq \frac{6}{5}w(t)+w(b)$. Since $a', c'$ are two edges of $opp(t)$ which do not contain a maximum weight edge $b'$, we get that $w(a')+2w(c')\leq w(opp(t)) \leq w(t)$. Also,  $w(c)-w(b)< \frac{2}{5}w(t) -\frac{1}{4}w(t)<\frac{1}{5}w(t)$.

\item  $\tilde C$ contains half-edges $(q_{in}, e^2_{pq}),  (r_{in}, e^2_{qr})$, i.e., $x=b$ and $y=c$.  We show that  $w(b)+w(c)+2w(c')\leq \frac{6}{5}w(t)+w(b)$, which is equivalent to $w(c)+2w(c') \leq \frac{6}{5}w(t)$. We know that $2w(c')<\frac{4}{5}w(t)$. On the other hand, $w(c)< \frac{2}{5}w(t)$.

\item $\tilde C$  contains also $(r_{in}, e^2_{pr}),  (q_{in}, e^2_{rq})$, i.e., $x=a'$ and $y=b'$.  We show that $w(a')+w(b')+2w(c')\leq \frac{6}{5}w(t)+w(b)$.
 Since $w(a')+w(b')+2w(c')=w(opp(t))+w(c') \leq w(t)+ \frac{2}{5}w(t)< \frac{7}{5}w(t)$ and $\frac{6}{5}w(t)+w(b)> (\frac{6}{5}+\frac{1}{4})w(t)=\frac{29}{20}w(t)$,
the inequality indeed holds.

\end{enumerate}
\koniec 

\koniec

{\bf Proof of Lemma \ref{uzup}}


Assume now that $c$ contains some chords.

There remains the question of how to ensure that each chord is safe. If a chord $e$ of $c$ is contained in a directed path $P$ consisting of edges of $C_{max}$ that contains some ray $r$ of $c$, then to guarantee that $e$ is safe, we can simply color all edges of $P$ between $e$ and the closest ray $r'$ of $c$ (together with $e$) with the same color as $r'$. 
 Assume then now that we have already colored all such chords of $c$. 

If $c$ has no rays, we color each chord either with $3$ or  $4$ and each edge of $c$ with both $1$ and $2$. For each subcycle $c'$ of $c$ we choose exactly one edge of $c$ and assign it color $4$, all the remaining edges of $c'$ are colored with $3$. We choose two edges of $c$ with $flex(e)=1$ and replace one one  $1$ with $3$ 
and on the other $2$ with $4$.

\koniec

{\bf Proof of Lemma \ref{tricky}} \\
\dowod  Let $M$ be a perfect matching of $G''$. If $M$ does not contain the edge $(a_{\{p,q,r\}}, b_{\{p,q,r\}})$ (which means that these vertices are matched via edges of weight $\frac{\kappa(t)}{2}$) and does not contain any edge corresponding to a half-edge of an edge of $c$, then $\hat C$ contains the loop $e'_t$
with weight $\kappa(t)$. If $M$ contains edges corresponding to all half-edges within $t$, then $\hat C$ contains all edges of $t$ as well as
the loop $e_t$ with weight $-\kappa(t)$. In other respects the proof is similar to that of Lemma \ref{relax}.\koniec

\end{document}